\documentclass[aps,10pt,pra,twocolumn,groupedaddress, floatfix, superscriptaddress,showpacs, showkeys,amsfonts]{revtex4-1}
\usepackage{amsfonts}
\usepackage{amsmath}
\usepackage{amssymb, setspace}
\usepackage{graphics,graphicx}
\graphicspath{{./}}
\usepackage{epsf}
\usepackage{subfigure}
\usepackage[%
  colorlinks=true,
  urlcolor=blue,
  linkcolor=blue,
  citecolor=blue
]{hyperref}
\usepackage[all]{hypcap}
\usepackage{color}
\usepackage[utf8]{inputenc}
\usepackage{dsfont}
\usepackage{enumitem}
\usepackage[utf8]{inputenc}

\usepackage{natbib}
\usepackage{braket}

\newcommand{\be}{\begin{equation}}
\newcommand{\ee}{\end{equation}}
\newcommand{\bea}{\begin{eqnarray}}
\newcommand{\eea}{\end{eqnarray}}
\newcommand{\eq}[1]{Eq.~\eqref{#1}}

\newcommand{\eqss}[2]{Eqs.~\eqref{#1}-\eqref{#2}}

\newcommand{\fig}[1]{Fig.~\ref{#1}}

\newcommand{\bem}{\begin{multline}}
\newcommand{\eem}{\end{multline}}

\newcommand{\ptr}[2]{{\rm Tr_{#1}}\left[#2\right]}

\newcommand\identity{1\kern-0.25em\text{l}}

\newcommand{\apporsm}[1]{App.~\ref{#1}}
\begin{document}

\title{Dissipative Dynamics of Graph-State Stabilizers with Superconducting Qubits}
\author{Liran Shirizly}\email{liran.shirizly@ibm.com}
\affiliation{IBM Quantum, IBM Research -- Israel, Haifa University Campus, Mount Carmel, Haifa 31905, Israel}
\author{Grégoire Misguich}\email{gregoire.misguich@ipht.fr}
\affiliation{Université Paris-Saclay, CNRS, CEA, Institut de Physique Théorique,\\
  91191 Gif-sur-Yvette, France}
\author{Haggai Landa}\email{haggaila@gmail.com}
\affiliation{IBM Quantum, IBM Research -- Israel, Haifa University Campus, Mount Carmel, Haifa 31905, Israel}

\begin{abstract}
We study experimentally and numerically the noisy evolution of multipartite entangled states, focusing on superconducting-qubit devices accessible via the cloud. We find that a valid modeling of the dynamics requires one to properly account for coherent frequency shifts, caused by stochastic charge-parity fluctuations. We introduce an approach modeling the charge-parity splitting using an extended Markovian environment. This approach is numerically scalable to tens of qubits, allowing us to simulate efficiently the dissipative dynamics of some large multiqubit states. Probing the continuous-time dynamics of increasingly larger and more complex initial states with up to 12 coupled qubits in a ring-graph state, we obtain a good agreement of the experiments and simulations. We show that the underlying many-body dynamics generate decays and revivals of stabilizers, which are used extensively in the context of quantum error correction. Furthermore, we demonstrate the mitigation of two-qubit coherent interactions (crosstalk) using tailored dynamical decoupling sequences. Our noise model and the numerical approach can be valuable to advance the understanding of error correction and mitigation and invite further investigations of their dynamics.

\end{abstract}

\maketitle
State-of-the-art qubit devices for quantum computation have been realized with tens and hundreds of qubits on single chips \cite{Osprey,arute2019quantum,moses2023race, wurtz2023aquila}. In many of those devices the models describing the control and environment errors are often similar, even when the underlying physical mechanisms are quite different. The noise sensitivity of the individual qubits and gate operations make quantum error correction codes an essential goal in the field, en route to harnessing the full power of quantum algorithms \cite{RevModPhys.87.307,sundaresan2023demonstrating, PhysRevX.11.041058,krinner2022realizing, postler2022demonstration,google2023suppressing, gupta2023encoding}.

Many quantum codes are based on storing information in delocalized, entangled $N$-qubit states ($N\gg 1$), and measuring $n$-qubit ($n$Q) operators (of low weight, $n\ll N$) for the detection of local errors and the application of corrections. A lot of effort is devoted to the development of numerical tools and characterization procedures, focusing both on the microscopic qubit dynamics and the high-level gates, and the question of whether the noise is Markovian (memoryless) or the contrary
\cite{bylander2011noise,PredictingNonMarkovian, Groszkowski2023simplemaster, Nielsen2021gatesettomography, PhysRevA.101.032343, lifshitz2021practical, puzzuoli2023algorithms, proctor2022scalable,green2022probing, santos2023scalable, agarwal2023modelling, wei2023characterizing}. In general, it is hard to model faithfully the interplay of various decoherence mechanisms and the continuous dynamics of coupled qubits. One of the outstanding challenges is the incorporation of noise parameters measured at the few-qubits level, in the regime of multiqubit-state dynamics.

In this Letter, we develop a fundamental noise model that is extensible to the many-body regime of qubit dynamics. We experimentally and numerically study the continuous-time dynamics of multiqubit graph states \cite{HEB04, hein_entanglement_2006}. Our experiments are conducted using \textsl{qiskit-experiments} \cite{kanazawa2023qiskit} on IBM Quantum superconducting transmon qubits accessible via the cloud \cite{IBMQuantum}. We characterize the 1Q and 2Q parameters relevant in the studied setup, together with state preparation and measurement (SPAM) errors. Identifying errors that may appear non-Markovian but can in fact be described using an appropriate Markovian environment, we employ a high-performance numerical solver \cite{lindbladmpo, LM23,fishman_itensor_2022} that allows us to efficiently handle the density matrix of many-qubit states. The simulation gives us access to state characteristics that are otherwise inaccessible.

\begin{figure}[t!]
    \centering
    \includegraphics[width=0.48\textwidth]{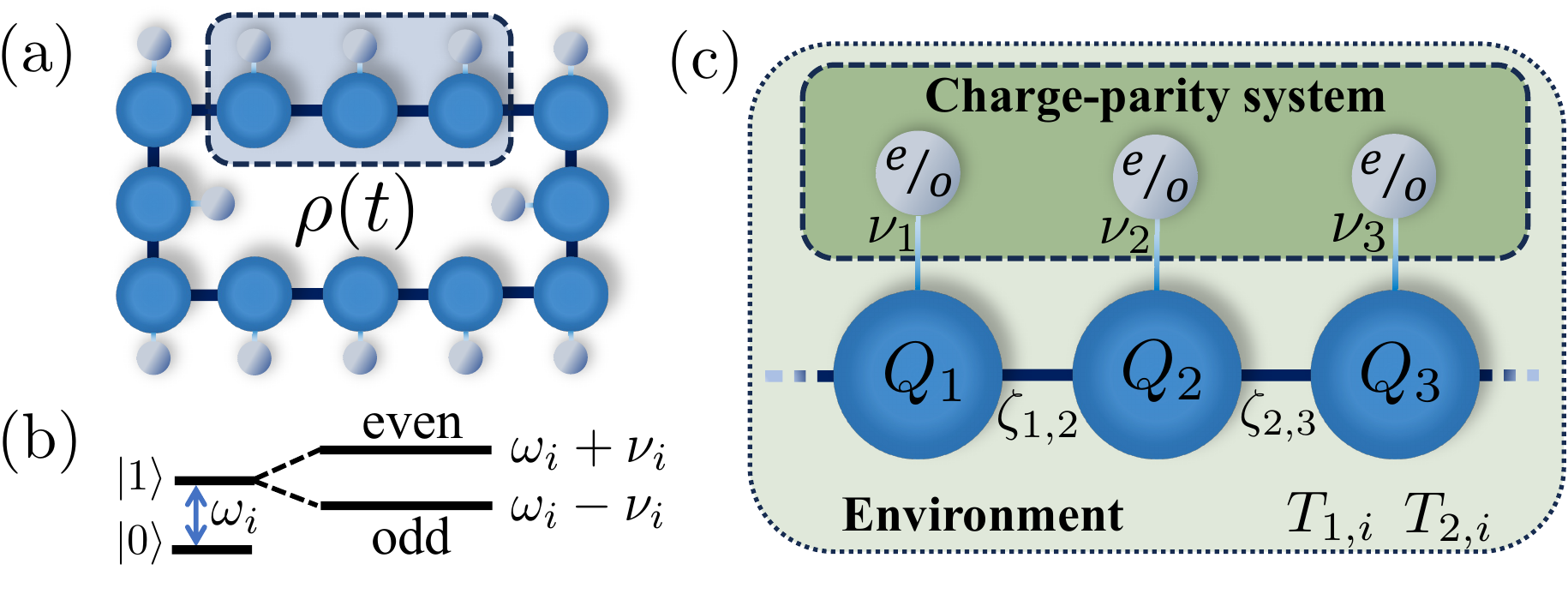}
    \caption{(a) This paper focuses on the dynamics of an open quantum system with a density matrix $\rho(t)$ of 12 qubits in a ring. (b) For superconducting qubits, the qubit levels (with frequency $\omega_i$) are split due to charge-parity fluctuations that manifest effectively as a Bernoulli stochastic variable shifting the qubit frequency by $\pm \nu_i$. (c) To model the charge-parity splitting in a many-body simulation reproducing the experiment dynamics, each qubit $Q_i$ is coupled to a fictitious two-level system [with levels denoted by $e$, $o$ (even/odd)] initialized to a diagonal mixed state, which is traced-over at the end of a calculation. The model further includes: energy relaxation time ($T_{1,i}$), dephasing time ($T_{2,i}$) and two-qubit ZZ crosstalk ($\zeta_{i,j}$) -- see the text for details.}
    \label{fig:schematic}
\end{figure}

Figure \ref{fig:schematic} shows a schematic depiction of the setup studied in this paper. The dynamical model that we consider applies generally to the decoherence of quantum memory states relevant for many physical systems. We start from the term in the dynamics specific to superconducting qubits, describing charge-parity oscillations. In essence, each qubit's frequency is shifted according to the charge parity (even or odd) of the qubit's junction electrodes, which switches due to quasiparticles tunneling. This splitting has been treated in the context of single-qubit experiments \cite{wilen2021correlated, PhysRevLett.121.157701,PhysRevB.84.064517,riste2013millisecond,PRXQuantum.3.030307,PhysRevB.100.140503, PhysRevLett.114.010501,10.21468/SciPostPhysLectNotes.31}, and in this work we present an approach for its inclusion as part of the basic noise model of many-body dynamics of superconducting qubits, essential for accurate simulations.

\begin{figure}[t!]
    \centering
    \includegraphics[width=0.45\textwidth]{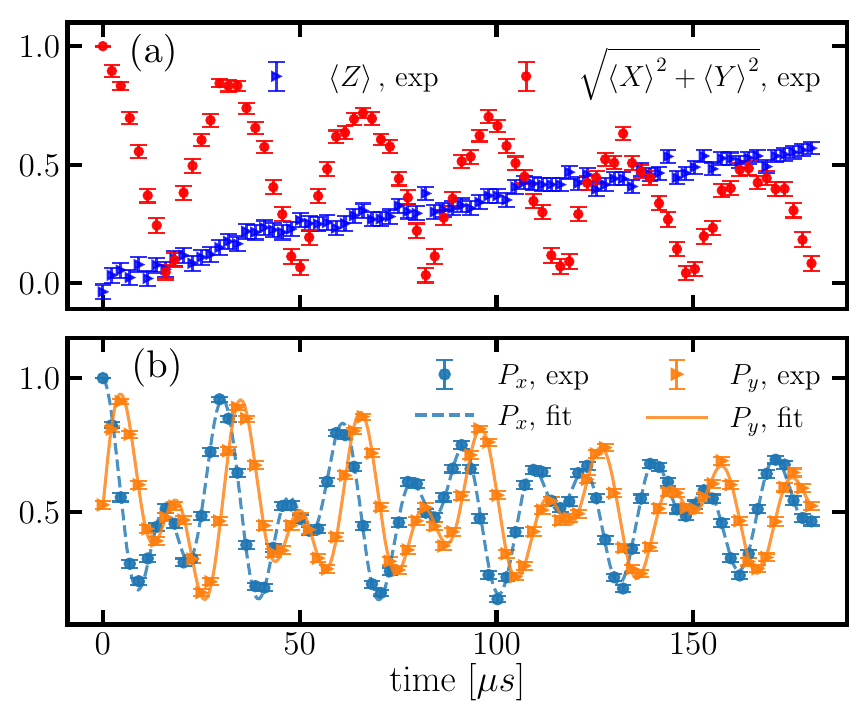}
    \caption{(a) A single qubit's mean  $xy$  projection of the Bloch vector ($\sqrt{\langle X\rangle^2+\langle Y\rangle^2}$) as a function of time after being initialized to the $|+\rangle$ state in a Ramsey experiment, plotted together with $\langle Z\rangle$ whose amplitude grows as the qubit's ground state becomes populated at a rate equal to $1/T_1$. The shrinking of the $xy$  projection and its revival is reminiscent of a non-Markovian process. (b) A characterization of $\nu$ and $\Delta$ of \eq{Eq:H_a} together with the decoherence time $T_2$ from the data points in an identical Ramsey experiment with the lines showing a fit of the data according to \eqss{Eq:P_x}{Eq:P_y}.}
    \label{fig:idle}
\end{figure}

As an example demonstrating the parity oscillations of a single qubit we consider a simple Ramsey experiment, wherein a qubit is prepared in the $|+\rangle=(|0\rangle + |1\rangle)/\sqrt{2}$ state (along the $+x$ direction of the Bloch sphere), and then its time evolution is probed with measurements in the $x$, $y$, and $z$ bases repeatedly to collect the probability of measuring the positive eigenstate. 
The magnitude of the qubit's Bloch vector projection on the $xy$ plane is plotted in \fig{fig:idle}(a) as a function of the time, together with $\langle z\rangle$. Here and in the rest of the paper, experimental data points and error bars indicate the mean and one standard deviation of 1024 measurements (shots) \footnote{The experiments were run on the device \emph{ibm\_cusco} between 07/20/2023 and 07/27/2023. Data from the experiments is in \cite{graph-state-dynamics}. Backends are listed in https://quantum-computing.ibm.com/}. The observed oscillations of the qubit's Bloch-vector norm could be assumed to result from an interaction (with a neighboring qubit or an uncontrolled degree of freedom) or a non-Markovian noise process, but this is in fact not the case here.

With superconducting qubit devices, each qubit's frequency is first characterized, to determine the microwave drive frequency to which each qubit is locked in experiments to follow. 
In the rotating frame with respect to this predetermined frequency, each of the parity states (denoted hereafter by a subscript $a\in \{e,o\}$) is subject to a Hamiltonian with a shifted frequency,
\begin{equation}
    H_a/\hbar = \frac{1}{2}\omega_a(1-\sigma^z),\qquad \omega_{e} = \Delta + \nu,\quad \omega_{o} = \Delta - \nu,\label{Eq:H_a}
\end{equation}
where $\Delta$ is the mean drift (or detuning) of the qubit's frequency from the microwave frame fixed previously, and $\nu$ the parity-oscillations frequency. Equation \eqref{Eq:H_a} adopts the convention that the qubit's ground state obeys $\sigma^z|0\rangle = |0\rangle$, while for the excited state: $\sigma^z|1\rangle = -|1\rangle$, and the higher levels beyond the first two are neglected.

The probabilities of even and odd parities have been taken in earlier experiments as being equal over appropriate timescales \cite{riste2013millisecond, PRXQuantum.3.030307,PhysRevB.100.140503, PhysRevLett.114.010501}. We test the consistency of this assumption with our model in the current device. Assuming that during each shot of the experiment the qubit's charge-parity is even or odd but constant, its density matrix $\rho(t)$ can be described as a convex sum of the independent parity contributions, $\rho = b\rho_e + (1-b)\rho_o$, where we introduce $b$ to parameterize the fraction of shots with even parity. By fitting $b$ as a free parameter we find that $b\approx 0.5$ almost always (within statistical noise), although we also find rare deviations (\apporsm{App:1QRamsey}). We set $b=1/2$ hereafter, describing well our data.

Fig.~\ref{fig:idle}(b) presents an example of a similar Ramsey experiment as described above, fitting the parameters of \eq{Eq:H_a} using the probabilities of measurements along the $x$ and $y$ directions. Each of the signals can be written as the product of two decaying oscillations (\apporsm{App:1QRamsey}),
\begin{align}
    P_x &= A \exp(-t/T_2)\cos[(\Delta+\omega_s) t + \phi] \cos(\nu t) + B, \label{Eq:P_x}\\
    P_y &= A \exp(-t/T_2)\sin[(\Delta+\omega_s) t + \phi] \cos(\nu t) + B, \label{Eq:P_y}
\end{align}
where $\omega_s$ is the ``intended'' frame detuning offset added to improve the signal, and $T_2$ is the dephasing time. In addition  $\phi$, $A$ and $B$ are fitting parameters accounting for the SPAM errors, which ideally would be 0, 1/2 and 1/2 respectively. We find that the model is consistent with the experiment data without requiring an additional Gaussian decay envelope corresponding to a $1/f$ noise \cite{krantz2019quantum} (\apporsm{App:GaussianDecay}). The characterized values of $\nu$ are consistent with the theoretical charge dispersion \cite{koch2007charge, schreier2008suppressing} of our used transmon qubits \apporsm{App:Stability}.

Once $\nu$ is characterized for each qubit, the  Hamiltonians $H_e$ and $H_o$ can be constructed numerically and the dynamics of the system can be simulated. However, an $N$-qubit simulation accounting for the charge-parity splitting would have to average the results of $2^N$ different time evolutions with their modified parameters corresponding to the initial conditions of even or odd parity. This quickly becomes intractable, and here we choose instead a different approach that allows us to scale our simulations to tens of qubits under relevant conditions.
For the purpose of simulation, we can map the problem of a qubit whose frequency is a (Bernoulli) random variable onto an open system with an additional fictitious ``qubit'' whose ground state labels the even parity, while its excited state relates to the odd parity. 
The Hamiltonian of the system can then be written as
\begin{equation}
    H_1 /\hbar = \frac{1}{2}\sum_{i} \left[ \Delta_i  + \nu_i\tilde{\sigma}^z_i\right]\left(1-\sigma^z_i\right),\label{Eq:H_1}
\end{equation}
where $\sigma^z_i$ is the Pauli $z$ matrix of the actual qubit, and $\tilde{\sigma}^z_i$ is the Pauli $z$ matrix corresponding to the parity.
The parity qubit can be described by a diagonal density matrix (parameterized with $b$), which naturally remains invariant under the time evolution.
 In this approach the system dimension apparently increases (exponentially) as compared with sampling of simulations with even/odd parameters. However, the Hamiltonian in \eq{Eq:H_1} is naturally suitable for a solver based on matrix product states (MPS) and matrix product operators (MPO), since the fictitious qubits do not develop entanglement with the system qubits, and only increase the computational requirements by a small amount.
 
In addition to the one-body Hamiltonian of \eq{Eq:H_1}, we run standard characterization experiments of the effective (approximate) ZZ interaction strength of every pair of idle qubits connected according to the device topology, taking the form \cite{PhysRevLett.129.060501, PhysRevA.101.052308}
\begin{equation}
    H_2 /\hbar =
    \frac{1}{2}\sum_{\left<i,j\right>}\zeta_{ij}
    \left(1-\sigma^z_i\right)\left(1-\sigma^z_j\right),\label{Eq:H_2}
\end{equation}
where the summation is over the nearest neighbors. The total Hamiltonian of the idle qubits is therefore $H=H_1+H_2$. To gain some understanding of the Hamiltonian dynamics, we consider the effect of tracing out all qubits except qubit $i$, which has $n_i$ nearest neighbors. The resulting 1Q density matrix evolution can be described as a mixture of $2^{1+n_i}$ effective qubits (see \apporsm{App:2QHamiltonian}), each oscillating coherently with different frequency $\omega_i \in \left\{\Delta_i \pm \nu_i + \sum_{j} \left(1\pm1\right) \zeta_{ij} \right\}$, where the sum is over the qubit's neighbors.

\begin{figure}[t!]
    \centering
    \includegraphics[width=0.47\textwidth]{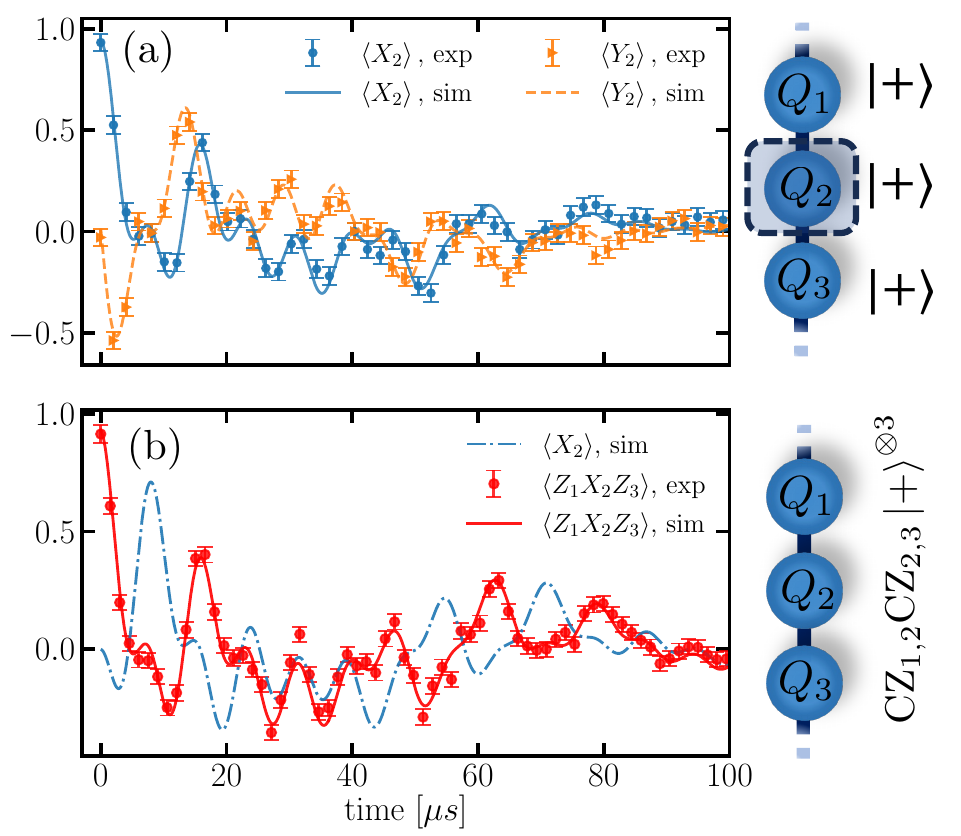}
    \caption{(a) Dynamics of the middle qubit among three that are initialized to the product state $|+\rangle^{\otimes 3}$. The experiment measurements are given by the points (with statistical error bars) and the lines are taken from simulation data. The multiple frequencies visible in the oscillations of the shown qubit result from the combination of its parity-oscillations, detuning error, and ZZ coupling to two neighbors (with different coupling strengths) -- see the text for a detailed discussion. (b) The dynamics of the $Z_1X_2Z_3$ stabilizer of a similar 3Q chain initialized in a graph state, and of the middle qubit's $\langle X\rangle$.
    In this figure and in \fig{fig:12Q}, the lines show simulation data, which once the Hamiltonian and noise parameters have been determined, do not involve any adjustable parameters.}
    \label{fig:3Q}
\end{figure}

Incorporating the dissipative dynamics is more complex \cite{sdid}, and
to capture the full dynamics evolved numerical tools are needed \cite{PhysRevLett.128.033602,LM23}. 
We solve a Lindblad master equation for $\rho(t)$, accounting for evolution with the Hamiltonian $H$ together with standard noise operators fed with the $T_1$ (lifetime due to spontaneous emission towards the ground state) and $T_2$ values of each qubit,
\be{\partial_t}\rho =  -\frac{i}{\hbar}[H,\rho]+\mathcal{D}[\sigma^+]+\mathcal{D}[\sigma^z]. \label{Eq:Lindblad}\ee
 In \eq{Eq:Lindblad} the dissipators take a standard form, $\mathcal{D}[\sigma^+]$ describes relaxation (spontaneous emission) towards each qubit's ground state, $\mathcal{D}[\sigma^z]$ describes dephasing (\apporsm{App:Master}). We have validated that a heating term can be neglected, since under typical conditions of superconducting qubits it is significantly suppressed \apporsm{App:Heating}.
 The initial state in the experiment is characterized accurately (self-consistently) and fed into the simulation, parameterized for each qubit by the three Bloch vector coordinates \cite{PhysRevResearch.4.013199}. Single-qubit readout errors are accounted for and mitigated (in the mean) in the experimental results by assuming uncorrelated errors, observed to be a very good approximation in current devices \cite{PhysRevResearch.4.013199, PRXQuantum.2.040326}. The continuous dynamics together with intermediate gates are solved with a high precision (see \apporsm{App:Master}).

In the rest of this paper we describe the results of experiments and simulations probing the dynamics of increasingly larger and more complex initial states. The parameters $\left\{\nu_i,\Delta_i,T_{1,i},T_{2,i},\zeta_{i,j} \right\}$, and SPAM parameters are determined by characterization experiments. We use the mean values of the estimated parameters for the simulations. The values are given in \apporsm{App:Values}.   
In \fig{fig:3Q}(a) we plot the time evolution of the middle qubit of three, initialized and simulated starting from the product state $|+\rangle^{\otimes 3}$.
The choice of an initial state in the equatorial plane of the Bloch sphere reveals the presence of multiple frequencies in the dynamics, visible in \fig{fig:3Q}(a).
Due to the ZZ coupling the qubits develop some entanglement while the competing incoherent processes damp the oscillations. The simulation captures this dynamics very precisely, and this is the result of focusing on data with a successful fitting of the parameters and the absence of drifts and jumps over the experiment duration, or interactions with uncontrolled degrees of freedom (see \apporsm{App:Stability}).

In the next step we perform a similar experiment and simulation, replacing the initial state by a three-qubit linear graph state, which is equivalent up to local rotations to a Greenberger-Horne-Zeilinger (GHZ) state \cite{GHZ_1989}, a maximally entangled state of three qubits. This graph state can be written explicitly as $|g\rangle = ({\rm CZ}_{1,2} {\rm CZ}_{2,3})|+\rangle^{\otimes 3}$, where ${\rm CZ}_{i,j}$ is the controlled-$Z$ gate applied to qubits $i$ and $j$. An $N$-qubit graph state can be characterized also as the unique eigenstate of all $N$ stabilizers with an eigenvalue 1, i.e. $S_k|g\rangle = |g\rangle$, where $S_k$ is a stabilizer of the graph state if it is the product of an $X$ operator on qubit $k$ and $Z$ operators on all of its neighbors in the graph \cite{HEB04}. As in quantum error correction, these stabilizers generate a commutative subgroup of the Pauli group that does not contain $-\identity$ \cite{gottesman1997stabilizer}.
In \fig{fig:3Q}(b) we present the dynamics of the  stabilizer $S_2 = Z_1 X_2 Z_3$ of the initial graph, where in this notation a capitalized letter from $\{X, Y, Z, I\}$ identifies a Pauli matrix or the identity, and the index indicates the qubit. The initial value of the stabilizer $\langle Z_1 X_2 Z_3\rangle$ in \fig{fig:3Q}(b) differs from 1 in our experiments due to preparation errors (see \apporsm{App:Preparation}). At intermediate times the stabilizer's oscillations are closely related to those of $\langle X_2\rangle$, which result from the combination of all Hamiltonian parameters as discussed above. 

\begin{figure}[t!]
    \centering
    \includegraphics[width=0.45\textwidth]{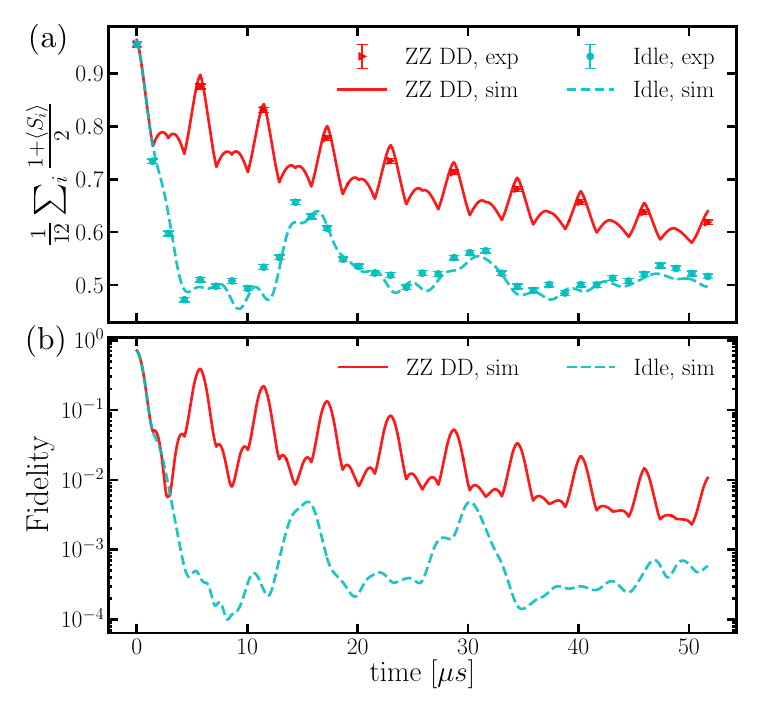}
    \caption{(a) The dynamics of the 12Q-ring graph state stabilizers' mean decoherence, as measured in the experiment (points) and extracted from simulations (lines). The presented measure that ideally equals 1 is shown with and without intermediate dynamical decoupling sequences, which cancel the effect of frequency shifts, charge-parity fluctuations, and ZZ coupling, see the text for details. (b) Using the data of the same simulations we can see how the fidelity of $\rho(t)$ with the ideal 12Q state is significantly improved when using the DD sequence canceling the coherent Hamiltonian errors.}
    \label{fig:12Q}
\end{figure}

We now turn to the largest setup studied in this work and our main result. We consider the dynamics of 12 qubits in a ring topology found in current IBM Quantum devices \cite{IBMQuantum,wootton2021hexagonal, PhysRevX.10.011022, doi:10.1126/sciadv.abi6690,hertzberg2021laser}, as depicted schematically in \fig{fig:schematic}(a). On such a ring, a translation-invariant graph state can be created efficiently using two layers of parallel CZ gates \apporsm{App:Preparation},
minimizing the initialization errors.
The 12 state stabilizers are $Z_{i-1}X_{i}Z_{i+1}$ where the $X_i$ operator is shifted along all qubits, and their expectation value can be measured using just two measurement setups, $X_1 Z_2 X_3 Z_4...X_{11}Z_{12}$ and $Z_1 X_2 Z_3 X_4...Z_{11}X_{12}$, and then tracing out the irrelevant qubits \cite{PhysRevResearch.2.043323}). This makes the (destructive) characterization of the state using its stabilizers very practical experimentally, and the relevance of the local stabilizers for characterizing complex states is well-motivated in the context of error correction codes. In \fig{fig:12Q}(a) we present a global measure of the deviation of the 12 stabilizers from their ideal expectation value of 1, derived by averaging over the positive quantities $(1+\langle S_i\rangle)/2$, giving the mean of the corresponding projection operators, to define $\bar{ P}=\frac{1}{N}\sum_i \frac{1}{2} ({1+\langle S_i\rangle})$. In the equilibrium (steady) state to which the system approaches for $t\gg T_{1,i}$ (which is close to the Hamiltonian ground state), we have $\langle Z_{i-1}X_{i}Z_{i+1} \rangle =0$ for all stabilizers. Therefore,  $\bar{ P}=1$ is the ideal case of  a perfect pure state, while in the ground state $\bar{ P}=1/2$. The presented experimental dynamics of $\bar{ P}$ are reproduced well in the simulation, and the individual 12 stabilizers are shown in \apporsm{App:12QStabilizers}. 

A natural next step is to consider the effect of dynamical decoupling (DD) in the current experiment. As follows from \eq{Eq:H_a} both the detuning (frame) errors and parity oscillations can be cancelled by standard 1Q DD sequences. The ZZ (crosstalk) interactions can be treated in parallel by staggering the single-qubit $X$ gates across the device according to a two-coloring of the interaction edges (a similar protocol including Y gates has been demonstrated in \cite{zhou2022quantum}). The total delay time $T_{\rm final}$ is sliced to $n_{\rm DD}=T_{\rm final}/T$ repetitions of DD sequences. Within each slice of delay time $T$, an X gate is applied on each of the qubits of the first colored sub-graph at times $T/2$ and $T$, and on the second sub-graph the X gates are applied at $T/4$ and $3T/4$ \apporsm{App:DD}. Figure \ref{fig:12Q}(a) shows the stabilizer dynamics obtained by adding this DD sequence. We measure the state at several different times $n_{\rm DD}\cdot T$ according to the number of DD repetitions $n_{\rm DD}=0,1,...,9$ with constant time slices $T$. Compared with the idle data, we measure less points in order to reduce the amount of error introduced into the experiment by the DD gates themselves (which can result, e.g., from gate inaccuracies, leakage of qubit wavefunctions out of the qubit manifold, and induced interactions). The improvement in $\bar{P}$ is clear and consistent. Although not all measured stabilizers agree exactly with the simulated ones (as can be seen by examining each of them in \apporsm{App:12QStabilizers}, where we discuss in more detail the discrepancies), the high degree of correspondence makes it plausible that the simulations capture the hardware dynamics to a large extent. This gives us a new powerful tool allowing us to calculate nonlocal quantities that are hard to track experimentally. As an example we take the data from the same simulations described above and show in \fig{fig:12Q}(b) the evolution of the full many-body fidelity of the noisy state with the ideal intended graph state. The fidelity is very sensitive to errors and the simulations indicate an improvement by about two orders of magnitude with the DD sequence.

To conclude, we have demonstrated the characterization of noisy dynamics of multipartite entangled states of superconducting qubits, together with a model and a numerical approach allowing for an accurate corresponding simulation. We find that the modelling of charge-parity oscillations is essential for a precise description of superconducting qubits. We emphasize that hardware dynamics often deviate from a Markovian model -- qubit parameters drift and fluctuate on various timescales and are subject to interactions with uncontrolled degrees of freedom \cite{carroll2022dynamics,hirasaki2023detection, PRXQuantum.4.020356}. In fact, the accuracy of the model in the presented cases is encouraging and could even be considered as surprising. We therefore consider this model as a first approximation that should constitute the fundamental dynamical model and be further elaborated.

The presented simulation method is scalable to tens of qubits in a Markovian environment, provided that the structure of entanglement in the simulated states is limited as imposed by typical tensor-network constraints. For example, GHZ states serve as a standard benchmark of quantum computers and can also be efficiently simulated with MPS, which holds for their dissipative dynamics as well \cite{houdayer2023solvable}. The Hamiltonian and dissipative dynamics in many qubit devices are similar to those we have considered or can be accounted for with some modifications \cite{cao2023generation, baumer2023efficient, graham2022multi,yang2022realizing,monz201114, monroe2021programmable}.
The realized graph state can be considered as a (simple) representative of a logical state of an error correction code \cite{cross2008codeword}. We show that the underlying many-body dynamics generate decays and revivals of the stabilizers, reflecting the different contributions of coherent versus incoherent error mechanisms and emphasizing the importance of properly modeling them.

Our entire experiment and simulation software is accessible as open source \cite{graph-state-dynamics}, and can be used as a starting point for a detailed study of qubit dynamics during quantum error correction protocols.

\section*{Acknowledgements}
We thank Yael Ben-Haim for contributions to the source code used in this research.
H.L. and L.S. thank Eli Arbel, Ted Thorbeck, Luke Govia, and Alexander Ivrii for very helpful feedback.
Research by H.L. and L.S. was sponsored by the Army Research Office and was accomplished under Grant Number W911NF-21-1-0002. The views and conclusions contained in this document are those of the authors and should not be interpreted as representing the official policies, either expressed or implied, of the Army Research Office or the U.S. Government. The U.S. Government is authorized to reproduce and distribute reprints for Government purposes notwithstanding any copyright notation herein. G.M. is supported by the PEPR integrated project EPiQ ANR-22-PETQ-0007 part of Plan France 2030.

\appendix
\widetext

\section{Charge-parity model and Ramsey experiment of superconducting qubit}\label{App:1QRamsey}
In this section we study the validity of the assumptions underlying our model of the charge-parity oscillations. Both the ground state and the excited state energy levels have a splitting, as well as the higher levels, and the value and sign of the splitting depend on the charge noise dispersion \cite{wilen2021correlated, PhysRevLett.121.157701,PhysRevB.84.064517,riste2013millisecond,PRXQuantum.3.030307,PhysRevB.100.140503, PhysRevLett.114.010501,10.21468/SciPostPhysLectNotes.31}. Since only the energy differences are relevant, when working with just the first two qubit levels the splitting can be absorbed into a single parameter $\nu$. In our notation, the labeling of even or odd is arbitrary and the frequency splitting $\nu$ is always taken to be positive.  

In our model, the qubit's density matrix $\rho(t)$ can be described as a convex sum of the independent parity contributions, \be \rho=b\rho_e+(1-b)\rho_o ,\label{Eq:rho_b}\ee where we introduce $b$ to parameterize the fraction of shots with even parity. Equation~\eqref{Eq:rho_b} is valid if the characteristic parity switching time (from even to odd or vice versa, without a change of the qubit's state) is large compared with a single experimental shot. This is justified \emph{a posteriori} since we see that our model agrees quantitatively within the error bars with the experimental data for many qubits. It is also consistent with earlier literature quoting a timescale of milliseconds for such a switching. Moreover, the IBM Quantum devices interleave the shots of all experiment circuits, thereby strongly mixing the even/odd occurrences distribution. 

To characterize the charge-parity oscillation parameters we use Ramsey experiments. In a Ramsey experiment, the qubit is initialized to the state $\ket{+}$ and after an idle time $t$, it is measured with respect to the x and y axes. In the ideal case without SPAM errors and charge-parity oscillations, the probabilities of measuring 0 along the x and y axis are, respectively,
\begin{equation}
\begin{split}
    P_{x,\text{ideal}} &= \frac{1}{2}e^{-t/T_2}\cos(\Delta t) + \frac{1}{2}, \\
    P_{y,\text{ideal}} &= \frac{1}{2}e^{-t/T_2}\sin(\Delta t) + \frac{1}{2},\label{Eq:Ramsey1}
\end{split}
\end{equation}
where $T_2$ is the dephasing time and $\Delta$ is the qubit's detuning. 

Summing over the charge-parity contributions with probabilities $P_x = b P_{x,e} + (1-b) P_{x,o}$ and $P_y = b P_{y,e} + (1-b) P_{y,o}$ gives, using \eq{Eq:Ramsey1},
\begin{equation}
\begin{split}
    P_x = \frac{1}{2}e^{-t/T_2}\left[ b \cos((\Delta + \nu) t) + (1-b)\cos((\Delta - \nu) t) \right]+ \frac{1}{2},\\
    P_y = \frac{1}{2}e^{-t/T_2}\left[ b \sin((\Delta + \nu) t) + (1-b) \sin((\Delta - \nu) t) \right]+ \frac{1}{2}.\label{Eq:Ramsey2}
\end{split}
\end{equation}

In order to get a better signal and a more sensitive fitting it is sometimes useful in Ramsey experiments to measure the qubit with respect to a known intended rotating frame that causes the signal to oscillate faster. This will shift the drifted detuning frequency $\Delta$ to $\Delta + \omega_s$, where $\omega_s$ is the intended known shift. 
In addition, because of SPAM errors, we add the to \eq{Eq:Ramsey2} further fitting parameters $A$, $B$ and $\phi$ which gives

\begin{equation}
\begin{split}
    P_x &= A e^{-t/T_2}\left[ b \cos((\Delta +\omega_s + \nu) t + \phi) + (1-b)\cos((\Delta +\omega_s - \nu) t + \phi) \right]+ B, \\
    P_y &= A e^{-t/T_2}\left[ b \sin((\Delta +\omega_s + \nu) t + \phi) + (1-b) \sin((\Delta  +\omega_s - \nu) t + \phi) \right]+ B. \label{Eq:Ram_app_with_b}
\end{split}
\end{equation}
If we set $b=1/2$, \eq{Eq:Ram_app_with_b} can be simplified using trigonometric identities to
\begin{equation}
\begin{split}
    P_x &= A \exp(-t/T_2)\cos[(\Delta+\omega_s) t + \phi] \cos(\nu t) + B, \\
    P_y &= A \exp(-t/T_2)\sin[(\Delta+\omega_s) t + \phi] \cos(\nu t) + B. \label{Eq:Ram_app}
\end{split}
\end{equation}

\begin{figure}[t!]
    \centering
    \includegraphics[width=0.45\textwidth]{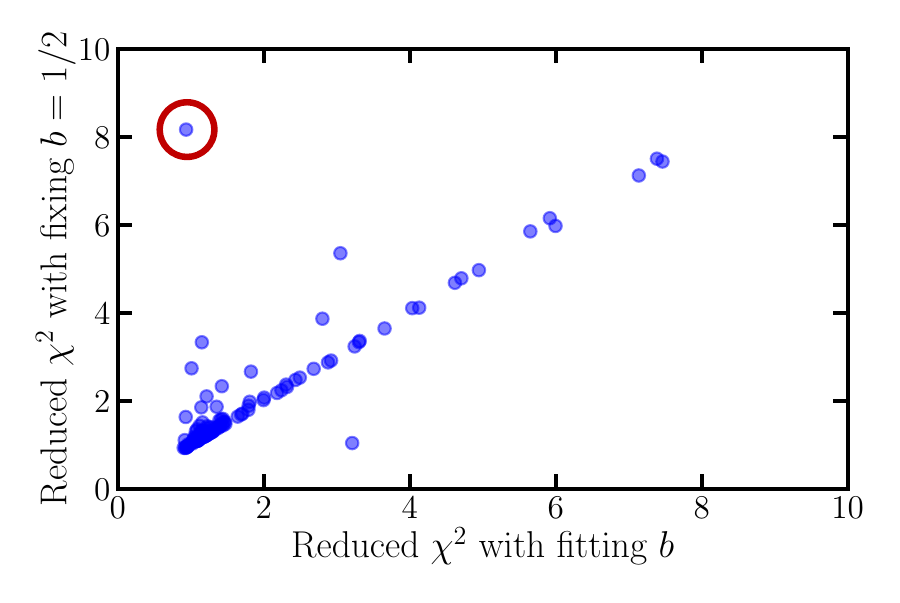}
    \caption{A comparison between the goodness of the fits with fixed parity imbalance $b$ and when $b$ is treated as a fit parameter, as quantified by their reduced-$\chi^2$ measure. Each point is given by fitting the same Ramsey characterization experiment data of one qubit from all of the qubits in \emph{ibm\_cusco} in one example realization. As shown, for most qubits setting $b=1/2$ is consistent with treating it as a free parameter. In \fig{fig:b_fit} we show full details about the main exception, marked here with a red circle.}  
    \label{fig:chi_b}
\end{figure}

\begin{figure}[h]
    \centering
    \includegraphics[width=0.9\textwidth]{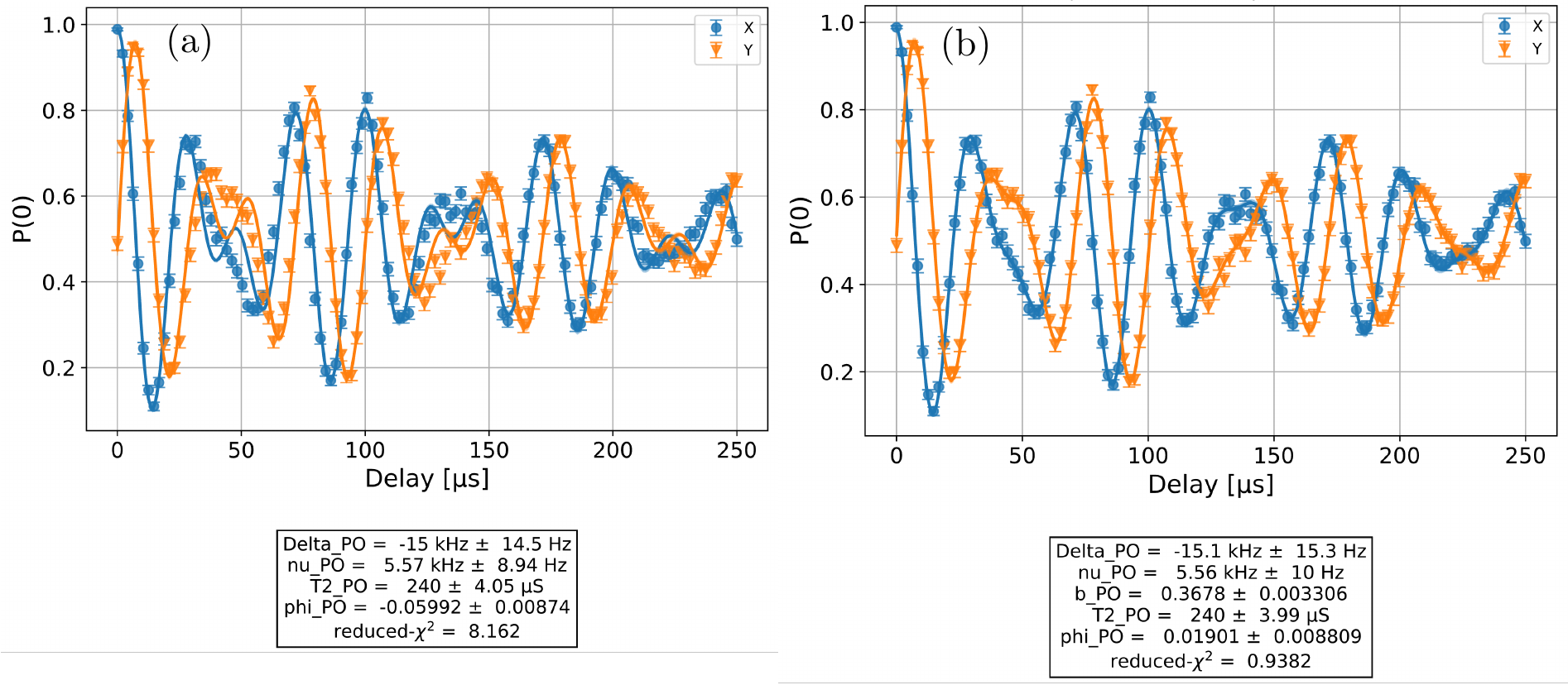}
    \caption{A Ramsey characterization experiment where a charge-parity fraction imbalance is strongly suggested by the data. (a) Experimental data (dots) and its fitted curve as in \eq{Eq:Ram_app} with equal parity probabilities ($b=1/2$). Clearly the fit doesn't capture the dynamics correctly (noticeable especially around $t\sim 50$), which can be also quantified by the high reduced-$\chi^2\approx 8.16$. (b) The same data fitted with the even parity fraction $b$ as a free parameter, as in \eq{Eq:Ram_app_with_b}. As shown this model captures the dynamics much better, with $b=0.368\pm0.003$ and a reduced-$\chi^2\approx 0.94$. The intended detuning is $\omega_s/2\pi=50$KHz.}  
    \label{fig:b_fit}
\end{figure}

In order to the test the necessity of the charge-parity imbalance parameter, we fit the experimental data of a control Ramsey experiment from all of \emph{ibm\_cusco}'s (127) qubits in one realization using the two models, \eq{Eq:Ram_app_with_b} and \eq{Eq:Ram_app}. As shown in \fig{fig:chi_b}, for most qubits setting $b=1/2$ is consistent with treating it as a free parameter. Therefore, we set $b=1/2$, describing well our data and use \eq{Eq:Ram_app} for characterization. We note that we do observe one main exception for this as shown in detail in \fig{fig:b_fit}, which clearly indicates a deviation. A possible speculative explanation could be that the parity switching time for this qubit is not much shorter than the overall time of the experiment, leading to a statistically observed bias in the initial states.

We treat this model as a simplified model of the full detailed dynamics of the current qubits, where of course there are exceptions, with qubits suspected to either couple to some uncontrolled degrees of freedom in the environment, manifest a charge jump during the experiment, or qubits subject to other unknown noise processes. The stability of the charge-parity splitting ($\nu$) is examined in the following, Sec.~\ref{App:Stability}. 

\section{Charge-parity stability}\label{App:Stability}

\begin{figure}[h]
    \centering
    \includegraphics[width=0.45\textwidth]{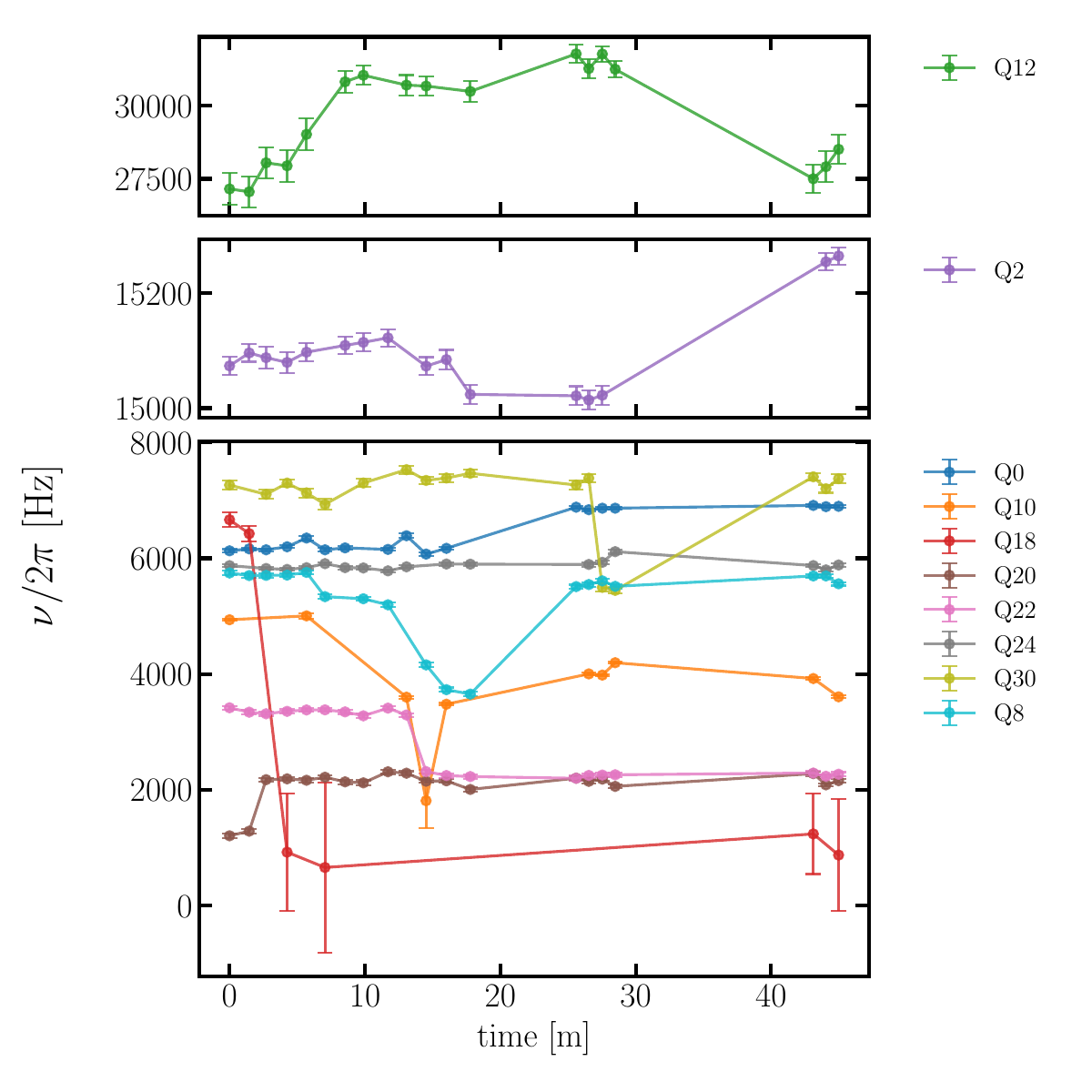}
    \caption{Charge-parity frequency splitting $\nu$, as a function of time for several qubits on  \emph{ibm\_cusco}. Each point is obtained from a different Ramsey experiment, the lines are added for clarity. Error bars indicate the uncertainty in the fitted values of $\nu$. The duration of each experiment is about 1 minute, and the time axis indicates the time that passed relative to the first experiment's completion time. The presented data shows typical intervals between jumps and typical ranges of $\nu$ values.}  
    \label{fig:nu_stability_cusco}
\end{figure}

\begin{figure}[h]
    \centering
    \includegraphics[width=0.45\textwidth]{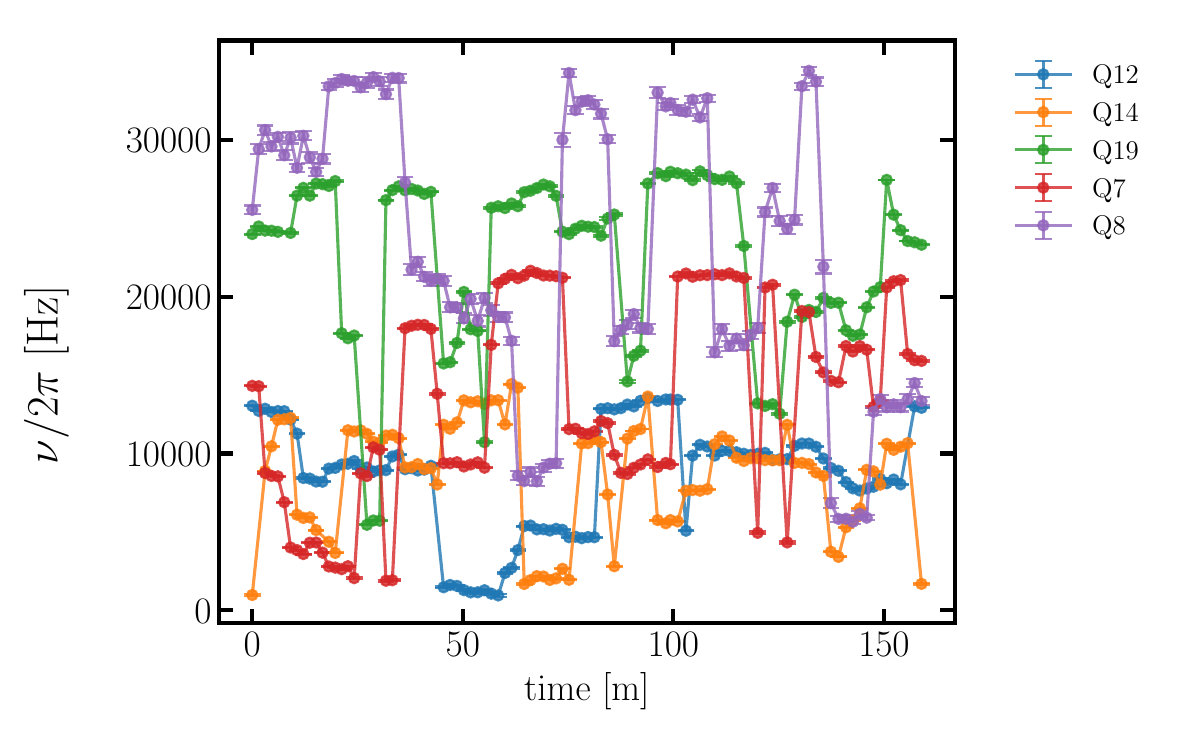}
    \caption{Charge-parity frequency splitting $\nu$, as a function of time for several qubits on a test Falcon device, similar to the ones available through \cite{IBMQuantum}. Each point is obtain from different Ramsey experiment, the lines are added for clarity. Error bars indicated the uncertainty in the fitted value of $\nu$.}  
    \label{fig:nu_sapporo}
\end{figure}

In this section we show the stability of the device charge-parity energy splitting. The stability of $\nu$ for several qubits is shown in \fig{fig:nu_stability_cusco} for the Eagle device (\emph{ibm\_cusco}) used in the experiments for this work. We see that for some qubits (e.g.,  Q22) the value is very stable for tens of minutes (with one jump within the presented interval of over 45 minutes), whereas for Q8 the value drifts more noteably. These values of $\nu$ are consistent with the theoretical charge dispersion \cite{koch2007charge, schreier2008suppressing} of our used transmon qubits, with median frequency of $\omega \approx5.12$GHz and median anharmonicity of $\alpha\approx-305$MHz. For comparison to other IBM Quantum devices, the stability with more data point on a test Falcon device is shown in \fig{fig:nu_sapporo}, where we see larger and more frequent jumps of $\nu$.

\section{Gaussian decay}\label{App:GaussianDecay}

\begin{figure}[h]
    \centering
    \includegraphics[width=0.45\textwidth]{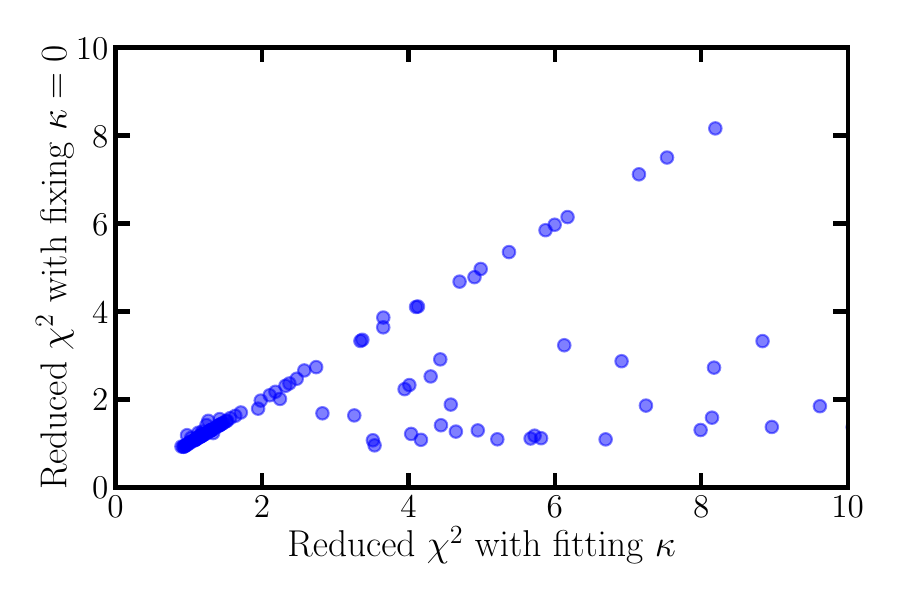}
    \caption{Compassion between the goodness of the fits with fixed Gaussian decay parameter $\kappa=0$ and fitting it, as quantify by their reduced $\chi^2$ measure. Each point is given by fitting the Ramsey characterization experiment data of a different qubit across \emph{ibm\_cusco} in one exampled realization. As shown, for all qubits, when setting $\kappa$ to zero, the fit model is at least as good as trying to fit it.}  
    \label{fig:kappa}
\end{figure}

\begin{figure}[h]
    \centering
    \includegraphics[width=0.75\textwidth]{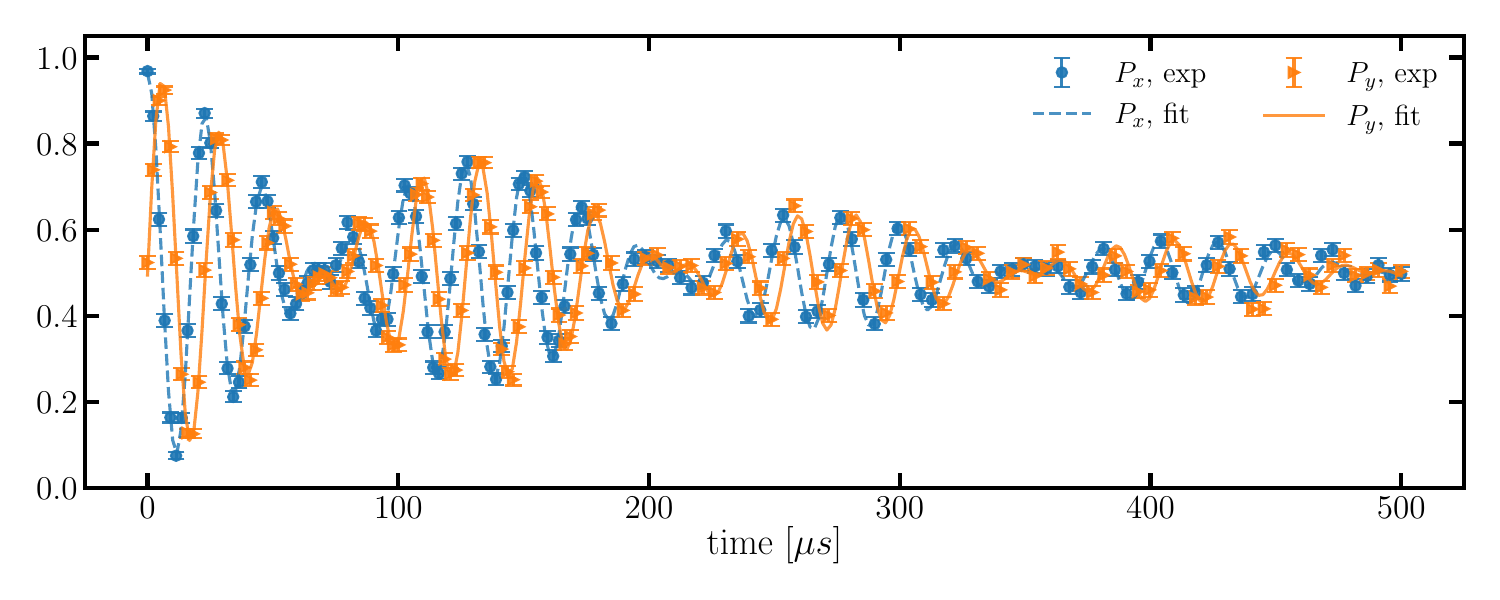}
    \caption{A Ramsey experiment with delays up to 0.5 milliseconds, with fitted curves as in \eq{Eq:Ram_app}. Such relatively long delays with high agreement to our model in many cases indicate the validity of the model.}  
    \label{fig:long_po}
\end{figure}

In some works on superconducting qubits, a Gaussian envelope appears in addition to the exponential decay of the Ramsey experiment signal \cite{krantz2019quantum}, which indicates the existence of noise with a $1/f$ power spectral density. In order to fit this effect we add an additional fit parameter $\kappa$, such that the Ramsey characterization become
\begin{align}
    P_x &= A \exp(-\kappa^2 t^2)\exp(-t/T_2)\cos[(\Delta+\omega_s) t + \phi] \cos(\nu t) + B, \\
    P_y &= A \exp(-\kappa^2 t^2)\exp(-t/T_2)\sin[(\Delta+\omega_s) t + \phi] \cos(\nu t) + B.
\end{align}
As shown in \fig{fig:kappa}, we find that $\kappa=0$ is consistent within our precision for all qubits that have been successfully fitted. We note that for $\nu t \ll 1$ the effect of the charge-parity oscillation is $\cos(\nu t)= 1 - \nu^2 t^2 /2$, which is the same as the leading-order contribution of some Gaussian envelope, therefore to bound the value of $\kappa$, longer delays are needed. For example we show in \fig{fig:long_po} an even longer Ramsey characterization with a very good fit (with $\kappa$ fixed 0) up to 0.5 millisecond.
Therefore for the qubits we used and the timescales we probed, a Gaussian decay envelope can be neglected.  

\section{Hamiltonian dynamics of two idle superconducting qubits}\label{App:2QHamiltonian}
In this section, the details of Hamiltonian dynamics are shown. Consider the Hamiltonian as described in the main text in the case of two qubits,
\be
H = \frac{1}{2}\left[ \Delta_1  + \nu_1\tilde{\sigma}^z_1\right]\left(1-\sigma^z_1\right) + \frac{1}{2}\left[ \Delta_2  + \nu_2\tilde{\sigma}^z_2\right]\left(1-\sigma^z_2\right) + \frac{1}{2}\zeta_{1,2}
    \left(1-\sigma^z_1\right)\left(1-\sigma^z_2\right).
\ee
Since the all the terms in the Hamiltonian are products of $\sigma_z$, it's eigenstates can written in the standard basis $\left\{ \ket{\sigma^z_1,\sigma^z_2}\otimes\ket{\tilde{\sigma}^z_1,\tilde{\sigma}^z_2} \right\}$, where we explicitly separate the parity qubits. Given the state \be \ket{\psi_0} = \left[ c_{00}\ket{0,0} + c_{10}\ket{1,0} + c_{01}\ket{0,1} + c_{11}\ket{1,1}\right] \otimes\ket{0,0}, \ee 
which has a defined (odd, odd) parity state, its Hamiltonian evolution is
\be
e^{-iHt}\ket{\psi_0} = \left[ c_{00}\ket{0,0} + e^{-i \left(\Delta_1 + \nu_1 \right) t}c_{10}\ket{1,0} + e^{-i \left(\Delta_2 + \nu_2 \right) t}c_{01}\ket{0,1} + e^{-i \left( \Delta_1 + \nu_1 + \Delta_2 + \nu_2 + 2\zeta_{1,2} \right)t}c_{11}\ket{1,1} \right] \otimes \ket{0,0}.
\ee
If we are interested only in the dynamics of one qubit, we need to trace out the second qubit. Tracing out the second qubit (and its parity) we get
\begin{equation}
\begin{split}
\rho_{Q1,o}(t) = &\ptr{{Q_2}}{e^{-iHt}\ket{\psi_0}\bra{\psi_0}e^{+iHt}} \\
= & \left( c_{00}\ket{0} + e^{-i \left( \Delta_1 + \nu_1 \right)t} c_{10}\ket{1} \right) \left(c_{00}^* \bra{0} + e^{+i \left( \Delta_1 + \nu_1 \right)t} c_{10}^* \bra{1}\right) + \\
&\left(c_{01}\ket{0} + e^{-i \left( \Delta_1 + \nu_1+ 2\zeta_{1,2} \right)t} c_{11}\ket{1}\right)\left( c_{01}^*\bra{0} + e^{+i \left( \Delta_1 + \nu_1+ 2\zeta_{1,2} \right)t} c_{11}^* \bra{1} \right).
\end{split}
\end{equation}
The full qubit state is a convex sum of the independent parity contributions. Repeating the derivation above for even parity gives the same result up to changing $\nu_1$ to $-\nu_1$. This gives the full dynamics of the first qubit: 
\be
\begin{split}\label{app:1q_Hamiltonion_dyn}
\rho_{Q_1}(t) =& \frac{1}{2}\left[\rho_{Q_1,o}(t) + \rho_{Q_1,e}(t) \right]  \\
=&\frac{1}{2} \left(c_{00}\ket{0} + e^{-i \left( \Delta_1 + \nu_1 \right)t} c_{10}\ket{1}\right)\left(c_{00}^* \bra{0} + e^{+i \left( \Delta_1 + \nu_1\right)t} c_{10}^* \bra{1}\right) + \\
&\frac{1}{2}\left(c_{01}\ket{0} + e^{-i \left( \Delta_1 + \nu_1+ 2\zeta_{1,2} \right)t} c_{11}\ket{1}\right)\left(c_{01}^*\bra{0} + e^{+i \left( \Delta_1 + \nu_1+ 2\zeta_{1,2} \right)t} c_{11}^* \bra{1}\right) +\\
&\frac{1}{2}\left(c_{00}\ket{0} + e^{-i \left( \Delta_1 - \nu_1 \right)t} c_{10}\ket{1}\right)\left(c_{00}^* \bra{0} + e^{+i \left( \Delta_1 - \nu_1\right)t} c_{10}^* \bra{1}\right) + \\
&\frac{1}{2}\left(c_{01}\ket{0} + e^{-i \left( \Delta_1 - \nu_1+ 2\zeta_{1,2} \right)t} c_{11}\ket{1}\right)\left(c_{01}^*\bra{0} + e^{+i \left( \Delta_1 - \nu_1+ 2\zeta_{1,2} \right)t} c_{11}^* \bra{1}\right) ,
\end{split}
\ee
which is basically a mixture of effective qubits $\rho_{Q_1}(t) = \sum_k c_k \rho_k(t)$, each oscillating with one of the frequencies: $\Delta_1 + \nu_1, \Delta_1 - \nu_1, \Delta_1 + \nu_1 + 2\zeta_{1,2}$ and $\Delta_1 - \nu_1 + 2\zeta_{1,2}$. 

Similarly, for qubit $i$ with $n_i$ nearest neighbors, its dynamics after tracing out the neighbors is a mixture of effective $ 2^{1+n_i}$ qubits (2 from its parity states, multiplied by 2 for each neighbor), \be \rho_{Q_i}(t) = \sum_k c_k \rho_k(t) \ee with frequencies:
\be
\omega_i \in \left\{\Delta_i \pm \nu_i + \sum_{j} \left(1\pm1\right) \zeta_{ij} \right\}. 
\ee

\section{Preparation of the ring graph state}\label{App:Preparation}

\begin{figure}[h]
    \centering
    \includegraphics[width=0.95\textwidth]{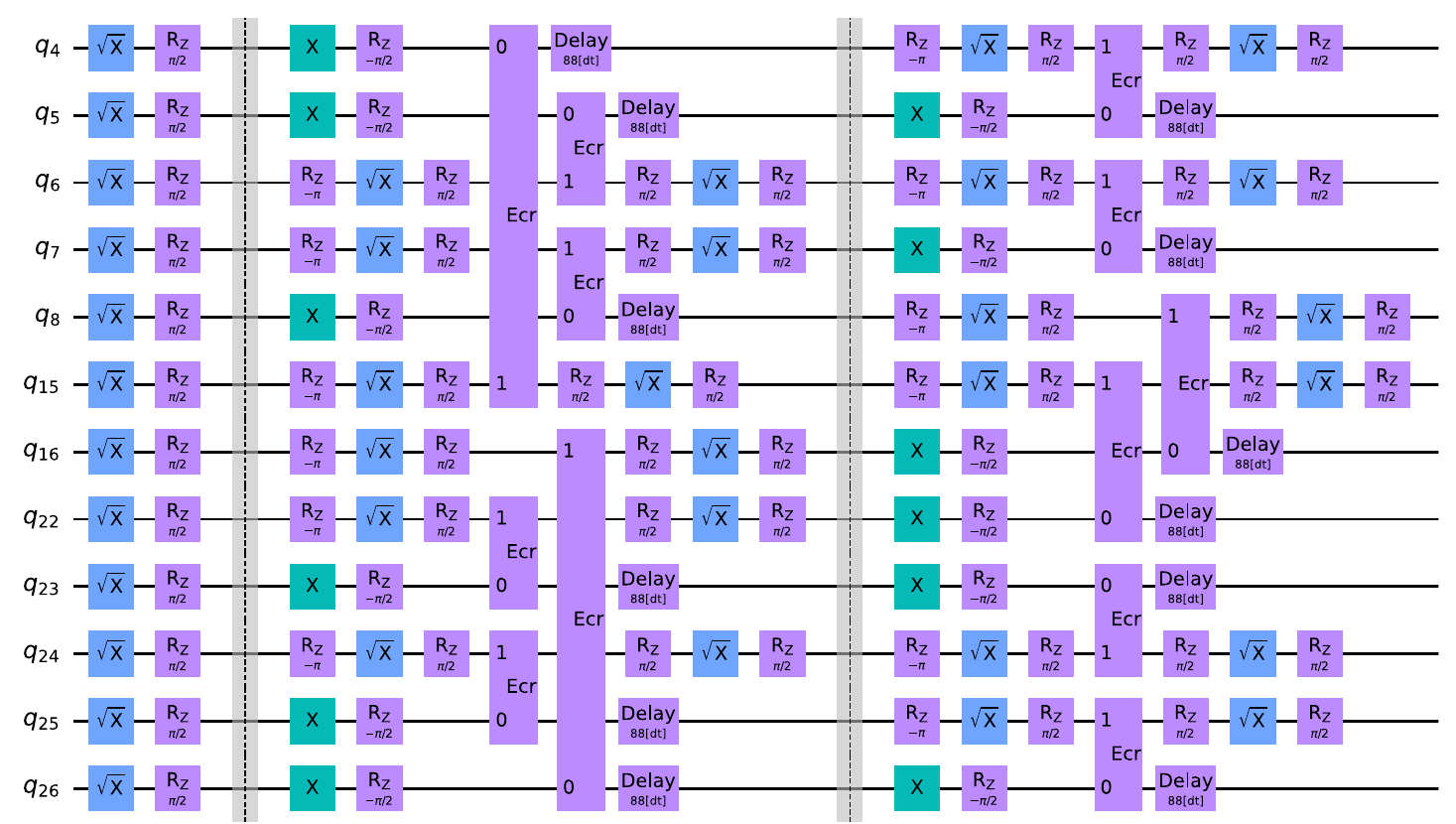}
    \caption{Quantum circuit initializing a 12-qubits ring graph state. The index shown for each qubit (in the notation $q_i$) is the physical qubit number actually used in the experiment. The first layer of circuits prepares the state $\ket{+}^{\otimes 12}$, then two layers of controlled-Z (CZ) gates follow according to the ring topology. Since CZ isn't a basis gate in \emph{ibm\_cusco}, equivalent transpiled circuits are implemented.}  
    \label{fig:graph_state_circuits}
\end{figure}

\begin{figure}[h]
    \centering
    \includegraphics[width=0.7\textwidth]{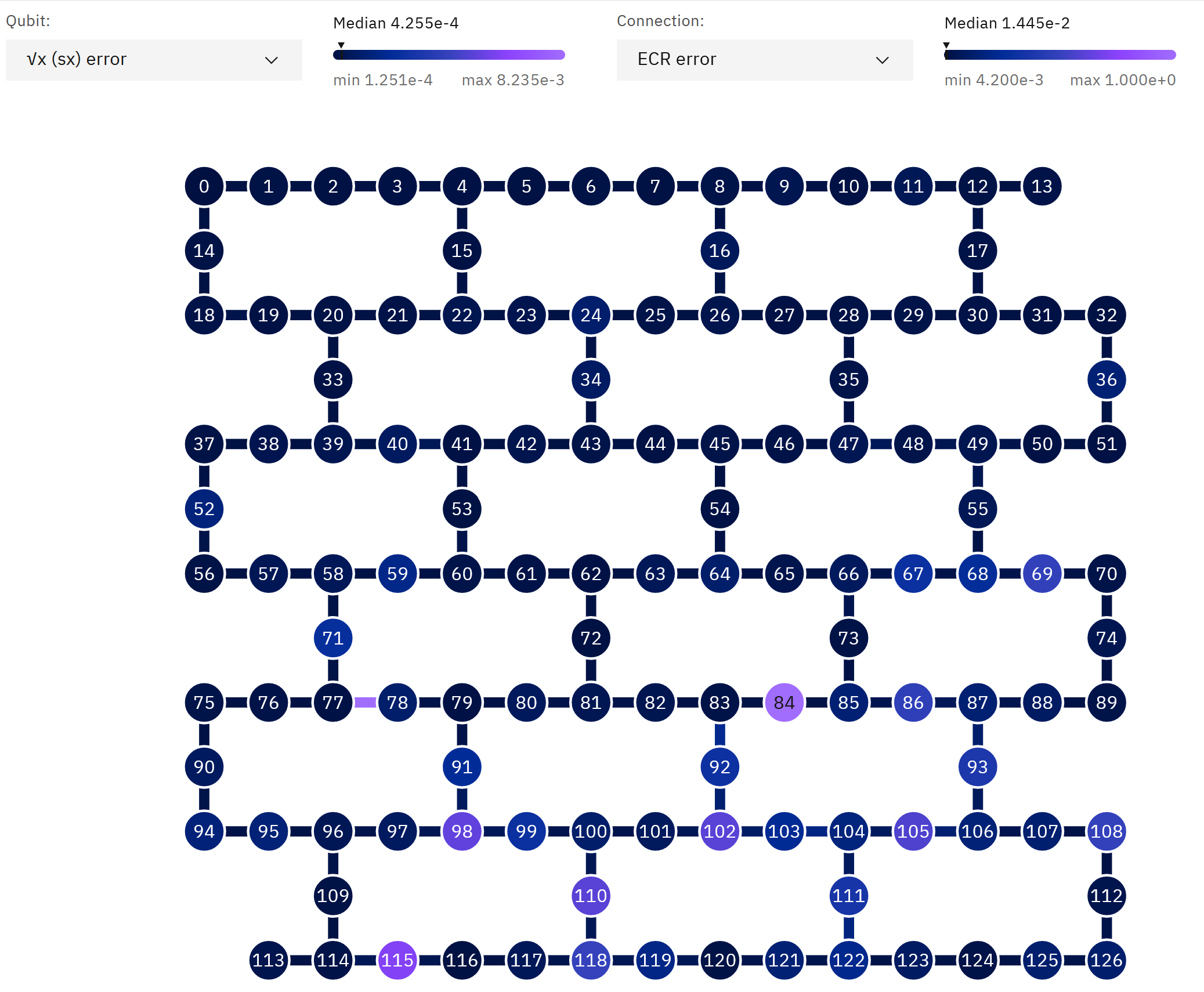}
    \caption{A schematic diagram of the device used in the experiments presented in this work (\emph{ibm\_cusco}, except one specific experiment shown in \fig{fig:nu_sapporo}). The distributions of 1Q and 2Q gate errors are depicted in the color map, and indicated at the top of the figure.}  
    \label{fig:Device}
\end{figure}

The ideal initial state is a 12-qubit ring graph state, which can be defined as
\be
\ket{g} = \text{CZ}_{12,1}\prod_{i=1}^{11}\text{CZ}_{i,i+1}\ket{+}^{\otimes 12}
\ee
where $\text{CZ}_{i,j}$ is the controlled-Z gate on qubits $i$ and $j$. The order of the CZ's is irrelevant since they commute. In the device we use the controlled-Z gate isn't a basis gate, so instead of each CZ, an ECR (echoed-cross resonance) gate with single-qubit rotations is applied. Two layers of ECR gates are required to generate a ring graph state, and the specific transpiled circuit we used is shown in \fig{fig:graph_state_circuits}. Due to the finite time of the gates and their errors the initialization of the graph state is imperfect. This is in addition to the noisy initial state, differing slightly from $\ket{0}^{\otimes 12}$ in the beginning of the circuit. The diagram of the entire device (\emph{ibm\_cusco}) is shown with in \fig{fig:Device} together with the distribution of single-qubit and two-qubit error values as measured in a specific calibration for illustration. The values of the device calibrations from the relevant experiment dates can be found in \cite{graph-state-dynamics}.

\section{Stabilizer dynamics of a twelve-qubit ring graph state}\label{App:12QStabilizers}

\begin{figure}[t!]
    \centering
    \includegraphics[width=0.99\textwidth]{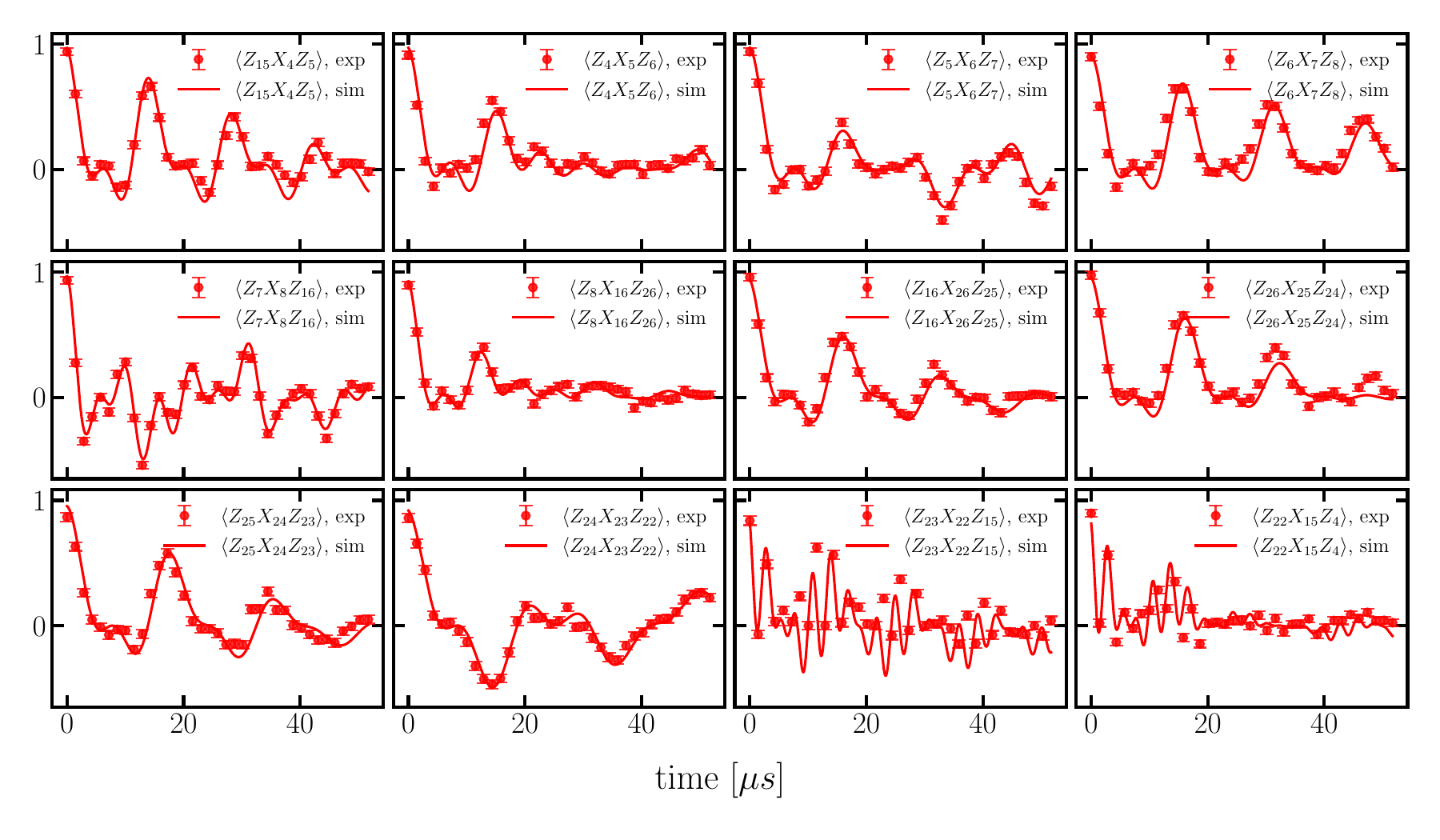}
    \caption{Dynamics of ring graph state's stabilizers without dynamical decoupling. The numbering are the physical qubit used for our experiment on \emph{ibm\_cusco}. Points are experimental data and lines are simulation. As shown, the simulation captures within error bars most of the experimental data. The main exceptions are stabilizers $\left< Z_{21}X_{22}Z_{15} \right>$, $\left< Z_{22}X_{15}Z_{4}\right>$ that oscillates faster then the others therefore more sensitive to the physical parameters.}
    \label{fig:all_stabilizers}
\end{figure}

\begin{figure}
    \centering
    \includegraphics[width=0.99\textwidth]{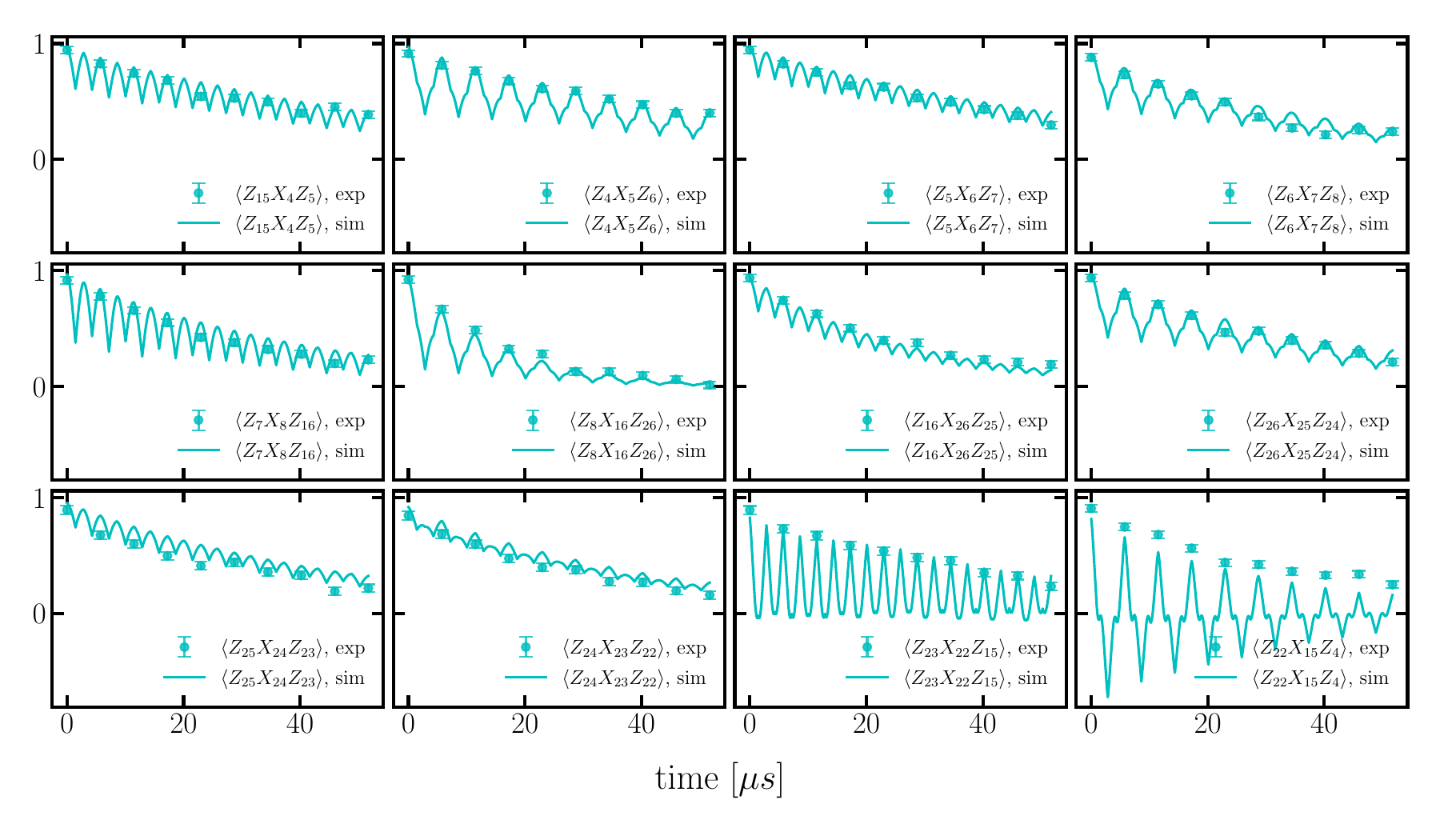}
    \caption{Dynamics of ring graph state's stabilizers with dynamical decoupling. The numbering are the physical qubit used for our experiment on \emph{ibm\_cusco}. Points are experimental data and lines are simulation. The time of application of the echoes of dynamical decoupling is according to the qubit location. On half of the ring (qubits: 5,7,16,25,23,15) the echoes are at $T/2$ and $T$ for a delay slice $T$, and on the other half (qubits: 4,6,8,26,24,22) at $T/4$ and $3T/4$. Two type of sharp changes in the dynamics of the stabilizers are shown according to the dynamical decoupling sequence of their middle qubit.}  
    \label{fig:all_stabilizers_DD}
\end{figure}

In this section, we show the dynamics of each one of the 12 stabilizers of the ring graph state. The dynamics of the stabilizers without dynamical decoupling is shown in \fig{fig:all_stabilizers}. We see a good agreement between the experimental data and our numerical solution for most of the stabilizers. The parameters used in the solver are taken from characterization experiments run before the stabilizers' evolution experiment and their values (including the simulation code) are available in \cite{graph-state-dynamics}. We used another characterization experiment at the end in order to test for jumps or drifts in the parameters. The last two stabilizers $\left< Z_{21}X_{22}Z_{15} \right>$, $\left< Z_{22}X_{15}Z_{4}\right>$ share one edge (between qubits 22 and 15) and oscillate more rapidly than the others. This is due to a ZZ crosstalk of magnitude 143kHz which is much higher than for the rest of the qubit pairs, which typically have values of a few tens of kHz. It seems that due to this large crosstalk, small variations between the characterization parameter values and their values during the stabilizer evolution experiment caused the numerics to deviate from the experiment.  

The dynamics of the graph state stabilizers including our dynamical decoupling sequences are shown in \fig{fig:all_stabilizers_DD}. The sharp changes are due to the fast echoes by the X gates. Our current model is in a good agreement to the experiment, even though it neglects gate errors and leakage from the qubit two levels. As in the previous figure, we see significant crosstalk between qubits 15 and 22, where rapid dynamics between the echoes (X gates) are shown. 

\section{Numerical master equation solution}\label{App:Master}

For the simulation we use the solver described in detail in \cite{LM23}, available as open source code in the public repository \cite{lindbladmpo}. The master equation for the density matrix $\rho(t)$ is
\be
\frac{\partial}{\partial t}\rho = -\frac{i}{\hbar}[H,\rho]+\mathcal{D}[\sigma^+]+\mathcal{D}[\sigma^z],\label{Eq:dtrho2}
\ee
where $[\cdot,\cdot]$ is the commutator of two operators, the Hamiltonian is given in the main text,
and the dissipators are defined by
\be \mathcal{D}[\sigma^+] = \sum_i g_{0,i}\left(\sigma_i^+ \rho\sigma_i^- - \frac{1}{2} \{\sigma_i^- \sigma_i^+,\rho\}\right),\label{Eq:D0}\ee
\be \mathcal{D}[\sigma^z] = \sum_i g_{2,i} \left(\sigma_i^z \rho\sigma_i^z - \rho\right),\label{Eq:D2}\ee
where the rates $g_0$ and $g_2$ for each qubit are related to the characteristic $T_1$ and $T_2$ times by
\be g_0 = 1/T_1,\qquad g_2=(1/T_2 - 1/2T_1)/2 ,\ee
[noting that following a common convention with superconducting qubits, the Hamiltonian has a negative sign in front of the qubit's $\sigma^z$ operator, and the relaxation (spontaneous emission) Lindblad operator is  $\sigma^+$].

A brute-force representation of the state of 12 qubits with their charge-parity environment and open system dynamics would require the memory equivalent of a Hamiltonian simulation of 48 qubits, which is not practical, while our simulation using an MPO representation of the density matrix and all operators in \eq{Eq:dtrho2} could be completed within a few hours on a laptop. 

We note one difference in the implementation of gates between the experiments and simulation. In the experiments the gates are implemented continuously and take a finite amount of time; $20 {\rm ns}$ for 1Q gates and $460 {\rm ns}$ for most qubits for the 2Q controlled-NOT ($CX$, implemented as ECR gates on the device \emph{ibm\_cusco}), which however are implemented instantaneously in the simulation. This approximation by itself is fine for the 1Q gates (less so for the 2Q gates which are significantly longer). In the current results there are either zero or two 2Q-gates involving each qubit during the initialization (and parallelized on all qubits in two layers), and hence for gates whose error per 2Q gate is in the range of about 0.01 (as measured in the device calibration data, downloaded from \cite{IBMQuantum} on the dates of the experiments, plotted for one instance in \fig{fig:Device} and available in \cite{graph-state-dynamics}), there is no notable difference with the simulation since the effect of two gates falls within the experiment error bars. Differences can be found with qubits on which the gates have relatively a low fidelity (due to various reasons), not modeled in the simulation, and due to accumulation of errors  during repeated DD sequences.

\section{Effect of qubit heating}\label{App:Heating}

In this section we study the effect of a qubit heating term that was neglected in \eq{Eq:dtrho2}, and verify that discarding this term is justified. The term which has been omitted is $\mathcal{D}[\sigma^-]$, which can be obtained from \eq{Eq:D0} by replacing $\sigma^+$ with $\sigma^-$ and vice versa. We denote the rate of this dissipator as $g_{1,i}$ (for qubit $i$). 

\begin{figure}
    \centering
    \includegraphics[width=0.5\textwidth]{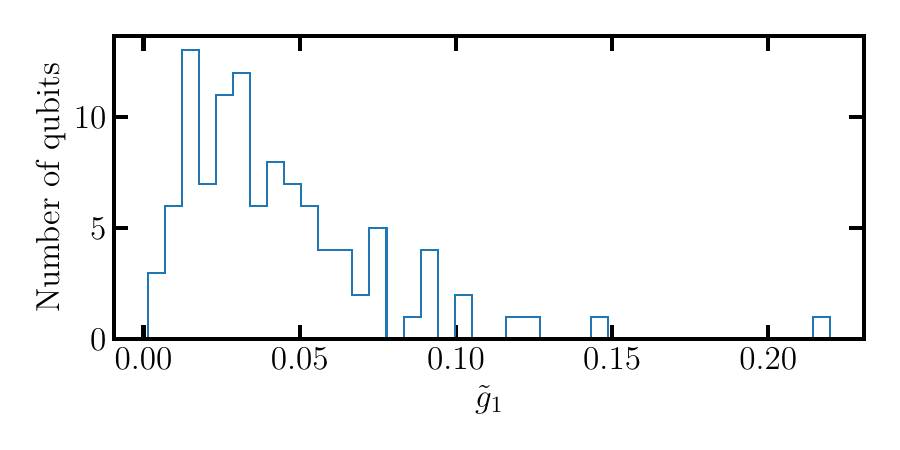}
    \caption{Distribution of the value $\tilde{g}_1$ across \emph{ibm\_cusco}, each value is estimated as in \eq{Eq:t1_decay}. The mean value is 0.043 and the median is 0.034. Bad qubits that could not be fitted are ignored.}   
    \label{fig:heating_term}
\end{figure}

We first run long-duration $T_1$ characterization experiment (inversion recovery) on all of \emph{ibm\_cusco}'s (127) qubits and fit each qubit's probability of measuring 1 (following readout mitigation) at varying durations according to the expression
\be \label{Eq:t1_decay} P(1) = A \exp(-t/T_1) +B.\ee
The fit parameters now relate to the master equation parameters by
\be 1/T_1 = g_0 + g_1, \qquad B = \frac{g_1}{g_0 + g_1}.\ee
Noting that $B$ has the physical meaning as the contribution fraction of the heating term to the overall decay, we rename it as the dimensionless ratio
\be \tilde g_1 = g_1 T_1 = B. \ee 
Figure \ref{fig:heating_term} presents the distribution of $\tilde{g}_1$, showing that the heating term is small compare to the overall decay across most qubits of \emph{ibm\_cusco}.     

\begin{figure}
    \centering
    \includegraphics[width=0.6\textwidth]{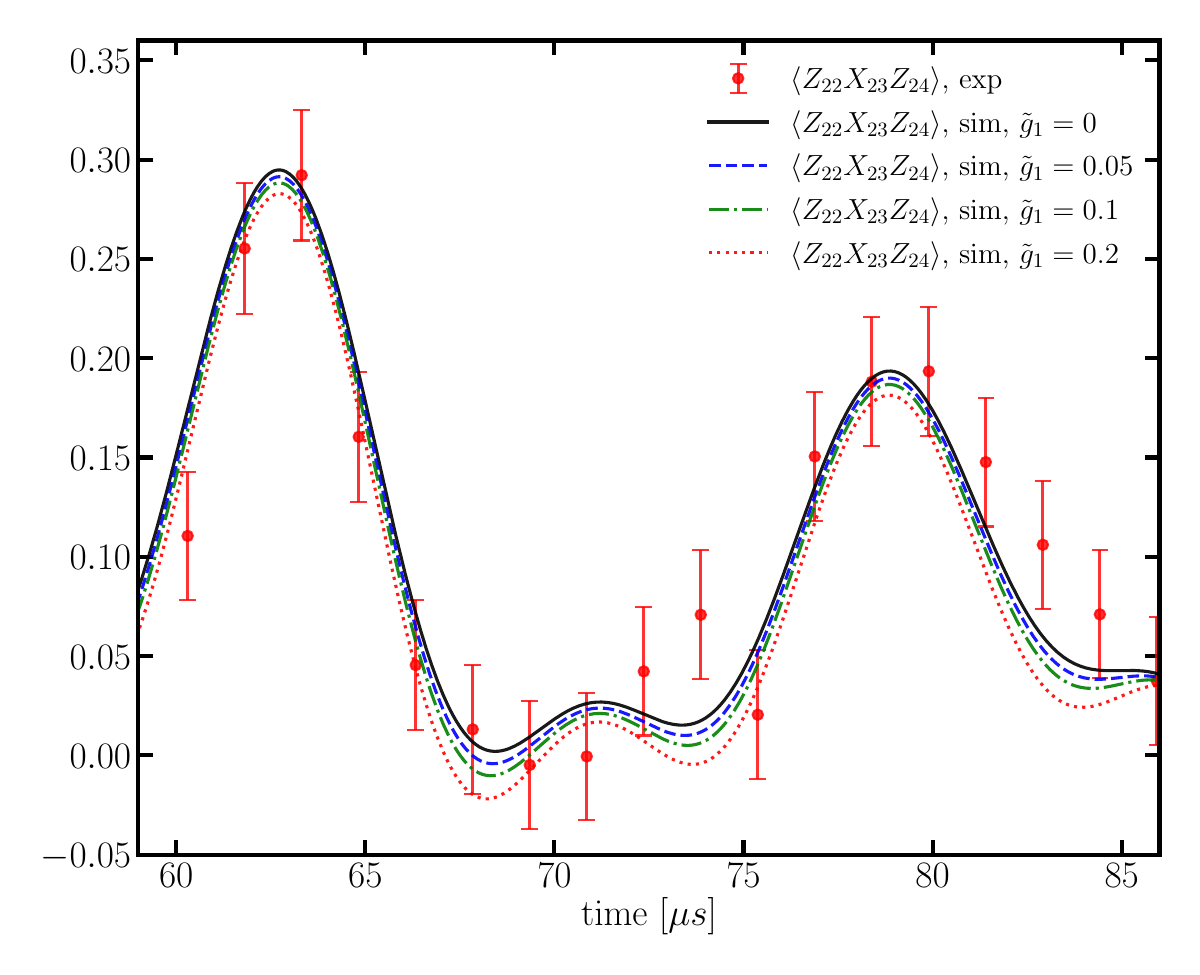}
    \caption{A zoom-in of the dynamics of the 3Q linear-chain graph state for increasing values of qubit heating rates, compared with the simulations with heating neglected and the experiment results. It is clearly seen that the heating effect at the examined values amounts to a small overall effect on the observables studied here, which falls well within the experiment error bars.}  
    \label{fig:heating}
\end{figure}

In order to examine the effect of the neglected $g_1$ rates in our simulations, we take the experiment data of the 3Q graph state of Figure 3(b) of the main text, and run three comparative simulations of the same dynamics with $g_1$ set to increasing values. We plot in \fig{fig:heating} simulations where each qubit's $g_1$ value has been set such that $\tilde g_1$ takes the values 0 (no heating), 0.05, 0.1, and 0.2. For  $\tilde g_1\neq 0$ this requires adjusting $g_0$ such that $T_1$ remains fixed at its experimentally characterized value of the original experiment. We see that the results of the simulations are almost indistinguishable (and well within the experiment error bars) even up to the excessively large value $\tilde g_2=0.2$, which is significantly more than typically observed with most qubits in the device when they are well behaved.

\clearpage
\section{Parameters values}\label{App:Values}
The values of the parameters $\left\{\nu_i,\Delta_i,T_{1,i},T_{2,i},\zeta_{i,j} \right\}$ determined in this work for each qubit are shown in the following tables: Table \ref{table:3Q_parameters+}, Table \ref{table:3Q_parameters_gr} and Table \ref{table:12Q_parameters}. In addition to these values, the full data and the code we used for running the experiments and simulations are given in \cite{graph-state-dynamics}.  

\begin{table}
\centering

$\begin{array}{|l |l |l |l |l||l | l|} \hline  
\text{Qubit} & T_1 \text{ [}\mu\text{s]}& T_2 \text{ [}\mu\text{s]}& \nu/2\pi \text{ [Hz]}& \Delta/2\pi \text{ [Hz]} & \text{Edge} & \zeta /2\pi \text{ [kHz]}\\ \hline \hline 
22 & 123\pm 4 & 84\pm2& 3308\pm33& -9088\pm50 & 22,23 & -39.4 \pm 0.4\\ \hline  
23 &  270\pm 21 &  104\pm3&  2253\pm30&  -12093\pm53 &23,24 &  -30.6 \pm 0.4\\ \hline  
24 &  249\pm 13 &  124\pm3&  5752 \pm 25 &  -6610 \pm 35 & &\\ \hline \hline

\end{array}$
\caption{Parameters values for the qubits used in the 3Q $\ket{+}^{\otimes3}$ state. In the main text, these qubits are labeled as qubits 1,2 and 3.}
\label{table:3Q_parameters+}
\end{table}

\begin{table}
\centering

$\begin{array}{|l |l |l |l |l||l | l|} \hline  
\text{Qubit} & T_1 \text{ [}\mu\text{s]}& T_2 \text{ [}\mu\text{s]}& \nu/2\pi \text{ [Hz]}& \Delta/2\pi \text{ [Hz]} & \text{Edge} & \zeta /2\pi \text{ [kHz]}\\ \hline \hline 
22 & 119\pm 4 & 81\pm2& 3277\pm34& -8345\pm50 & 22,23 & -39.1 \pm 0.4\\ \hline  
23 &  252\pm 2 &  114\pm5&  1054\pm66&  -13795\pm 38 &23,24 &  -31.1 \pm 0.4\\ \hline  
24 &  200\pm 10 &  104\pm3&  5570 \pm 28 &  -6910 \pm 41 & &\\ \hline 

\end{array}$
\caption{Parameters values for the qubits used in the 3Q chain graph state. In the main text, these qubits are labeled as qubits 1,2 and 3.}
\label{table:3Q_parameters_gr}
\end{table}

\begin{table}
\centering

$\begin{array}{|l |l |l |l |l||l | l|} \hline  
\text{Qubit} & T_1 \text{ [}\mu\text{s]}& T_2 \text{ [}\mu\text{s]}& \nu/2\pi \text{ [Hz]}& \Delta/2\pi \text{ [Hz]} & \text{Edge} & \zeta /2\pi \text{ [kHz]}\\ \hline \hline 
4 & 155\pm 5 & 148\pm4& 2269\pm19& -4867\pm35 & 4,5 & -38.4 \pm 0.3\\ \hline  
5 &  172\pm 6 &  146\pm3&  6252\pm22&  -6153\pm28 &5,6 &  -29.5 \pm 0.3\\ \hline  
6 &  250\pm 13 &  183\pm4&  11087 \pm 19 &  -4525 \pm 28 & 6,7 & -32.9 \pm 0.3\\ \hline  
7 &  172\pm7&  142\pm6&  1254 \pm 37 &  -858 \pm 35 & 7,8 & -31.1 \pm 0.4 \\ \hline  
8 &  79\pm2&  76\pm2&  2200 \pm 44 &  36103 \pm 62 & 8,16 & -45.8 \pm 0.4 \\ \hline  
16 &  163\pm6&  224\pm8&  2832 \pm 353 &  -1400 \pm 190 & 16,26 & -32.9 \pm 0.3 \\ \hline  
26 &  157\pm5&  42\pm1&  3199 \pm 83 &  -6127 \pm 107 & 26,25 & -31.7 \pm 0.3 \\ \hline  
25 &  132\pm4&  99\pm3&  5041 \pm 27 & -1071 \pm 45 & 25,24 & -29.8 \pm 0.4  \\ \hline  
24 &  173\pm7&  110\pm3&  3890 \pm 27 &  -6428 \pm 40 & 24,23 & -30.6 \pm 0.4  \\ \hline  
23 &  221\pm13&  111\pm4&  2204 \pm 30 &  -40798 \pm 52 & 23,22 & -38.8 \pm 0.4 \\ \hline  
22 &  100\pm2&  93\pm2&  3126 \pm 30 & -6368 \pm 45 & 22,15 & -142.6 \pm 0.4 \\ \hline  
15 &  229\pm12& 48\pm1 & 7909\pm60 & 6708\pm76 & 15,4 & -34.8 \pm 0.3 \\ \hline 

\end{array}$
\caption{Parameters values for the qubits used in the 12Q ring graph state.}
\label{table:12Q_parameters}
\end{table}

\clearpage
\section{Dynamical decoupling}\label{App:DD}
\begin{figure}[h]
    \centering
    \includegraphics[width=0.4\textwidth]{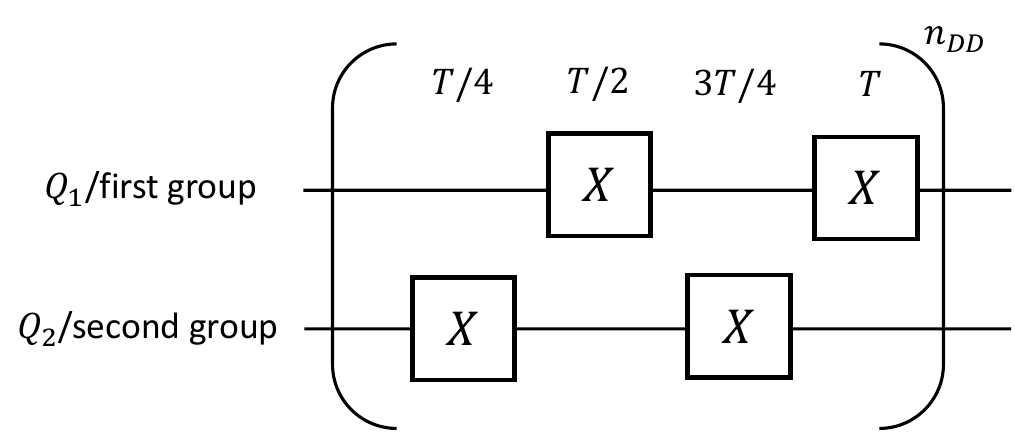}
    \caption{Staggered dynamical decoupling sequence. A total delay time $T_{\rm final}$, is sliced to $n_{\rm DD}=T_{\rm final}/T$ repetition of the following DD sequence: X gates are applied on the first qubit (or first group of qubits) at $T/2$ and $T$, and X gates is applied on the second qubit (or the second group of qubits) at $T/4$ and $3T/4$.}  
    \label{fig:DD_seq}
\end{figure}

In this section we will give more details about the DD sequences we used. In order to mitigate the ZZ cross-talk in addition to the single qubits terms, we used a staggered DD sequences. We will first describe it for simplified case of two connected qubits with shared edge. A total delay time $T_{\rm final}$, is sliced to $n_{\rm DD}=T_{\rm final}/T$ repetition of the following DD sequence: X gates are applied on the first qubit at $T/2$ and $T$, and X gates is applied on the second qubit at $T/4$ and $3T/4$ as shown in \fig{fig:DD_seq}. That way, the phase accumulation from the ZZ cross-talk is distributed equally to positive and negative. That is as opposed to common DD, where all the echoes applied simultaneously, and the sign of the ZZ cross-talk stays the same. This idea can be straight forward generalized to any group of qubits with topology of two-coloring graph, which is the case for our ring topology. For two-coloring graph, the gates are applied on the first (second) group as on the first (second) qubit.

In our experiments we compare the idle dynamic to the dynamics of DD version, we choose to measure the state in several different times according to the number of DD repetitions $n_{\rm DD}=0,1,...,9$ with constant time slice $T=5.757\mu s$. That means that for given $n_{\rm DD}$, we applied X gates on the first group of qubits (physical qubits: 5,7,16,25,23,15) at $T/2, T, 3T/2, ...,n_{\rm DD}\cdot T$ and on the second group (physical qubits: 4,6,8,26,24,22) at $T/4, 3T/4, 5T/4, ..., \left(4n_{\rm DD}-1\right)T/4$.

%\clearpage

%\bibliographystyle{plain}
\bibliography{references}

\end{document}